\documentclass[a4paper,fleqn]{cas-sc}

\usepackage{lineno}
\newcommand*\patchAmsMathEnvironmentForLineno[1]{
  \expandafter\let\csname old#1\expandafter\endcsname\csname #1\endcsname
  \expandafter\let\csname oldend#1\expandafter\endcsname\csname end#1\endcsname
  \renewenvironment{#1}
     {\linenomath\csname old#1\endcsname}
     {\csname oldend#1\endcsname\endlinenomath}}
\newcommand*\patchBothAmsMathEnvironmentsForLineno[1]{
  \patchAmsMathEnvironmentForLineno{#1}
  \patchAmsMathEnvironmentForLineno{#1*}}
\AtBeginDocument{
\patchBothAmsMathEnvironmentsForLineno{equation}
\patchBothAmsMathEnvironmentsForLineno{align}
\patchBothAmsMathEnvironmentsForLineno{flalign}
\patchBothAmsMathEnvironmentsForLineno{alignat}
\patchBothAmsMathEnvironmentsForLineno{gather}
\patchBothAmsMathEnvironmentsForLineno{multline}
}

\usepackage[authoryear,longnamesfirst]{natbib}
\usepackage{physics}

\def\tsc#1{\csdef{#1}{\textsc{\lowercase{#1}}\xspace}}
\tsc{WGM}
\tsc{QE}

\begin{document}

\let\WriteBookmarks\relax
\def\floatpagepagefraction{1}
\def\textpagefraction{.001}

\shorttitle{Global N-body Simulation of Gap Edge Structures in Saturn's Rings}    

\shortauthors{Torii et al.}  

\title[mode = title]{Global N-body Simulation of Gap Edge Structures Created by Perturbations from a Small Satellite Embedded in Saturn's Rings}

\author[1,2]{Naoya Torii}[orcid=0009-0003-5452-7473]
\ead{torii.n.aa@m.titech.ac.jp}
\affiliation[2]{organization={Department of Earth and Planetary Sciences, Tokyo Institute of Technology},
            addressline={Ookayama}, 
            city={Meguro-ku, Tokyo},
            postcode={152-8551}, 
            country={Japan}}

\affiliation[2]{organization={Earth-Life Science Institute, Tokyo Institute of Technology},
            addressline={Ookayama}, 
            city={Meguro-ku, Tokyo},
            postcode={152-8550}, 
            country={Japan}}

\author[1,2]{Shigeru Ida}
\ead{ida@elsi.jp}
\affiliation[3]{organization={National Astronomical Observatory of Japan},
            addressline={Osawa}, 
            city={Mitaka, Tokyo},
            postcode={181-8588}, 
            country={Japan}}

\author[3]{Eiichiro Kokubo}
\ead{kokubo.eiichiro@nao.ac.jp}
\affiliation[4]{organization={Department of the Study of Contemporary Society, Kyoto Women's University},
            addressline={Imakumano}, 
            city={Higashiyama, Kyoto},
            postcode={605-8501}, 
            country={Japan}}

\author[4]{Shugo Michikoshi}
\ead{michikos@kyoto-wu.ac.jp}

\begin{abstract}
Small satellites embedded in Saturn's ring - Daphnis and Pan - open a clear gap. 
Observations by the Voyager and Cassini spacecrafts have revealed various striking features of the gap structure, such as the density waves, sharp edge, and vertical wall structure. 
In order to explain these features in a single simulation, we perform a high-resolution ($N\sim10^6-10^7$) global full N-body simulation of gap formation by an embedded satellite considering gravitational interactions and inelastic collisions among all ring particles and the satellite, while these features have been mostly investigated separately with different theoretical approaches: the streamline models, 1D diffusion models, and local N-body simulation.
As a first attempt of a series of papers, we here focus on the gap formation by separating satellite migration with fixing the satellite orbit in a Keplerian circular orbit. 
We reveal how the striking gap features - the density waves, sharp edge, and vertical wall structure - are simultaneously formed by an interplay of the satellite-ring and ring particle-particle interactions.
In particular, we propose a new mechanism to quantitatively explain the creation of the vertical wall structure at the gap edge.
Inelastic collisions between ring particles damp their eccentricity excited by the satellite's perturbations to enhance the surface density at the gap edge, making its sharp edges more pronounced.
We find the eccentricity damping process inevitably raises the vertical wall structures the most effectively in the second epicycle waves.
Particle-particle collisions generally convert their lateral epicyclic motion into vertical motion. 
Because the excited epicyclic motion is the greatest near the ring edge and the epicycle motions are aligned in the first waves, the conversion is the most efficient in the gap edge of the second waves and the wall height is scaled by the satellite Hill radius, which are consistent with the observations.
\end{abstract}


\maketitle

\section{Introduction}\label{Introduction}
Saturn's ring is mainly composed of a large number of icy particles, ranging in size from a few centimeters to a few meters \citep{Cuzzi:2018}. 
Several small satellites are embedded in the A ring, ranging in size from a few hundred meters to a few kilometers. 
These small satellites gravitationally interact with the surrounding ring particles and create remarkable structures, which have been discovered by Voyager and Cassini spacecrafts' observations, such as propellers \citep{Tiscareno:2006}, gaps with satellite wakes and density waves \citep{Cuzzi:1985, Showalter:1991, Porco:2005}, and mountain-like vertical walls \citep{Weiss:2009}.
A satellite embedded in a planetary ring tends to open a gap by its gravitational perturbations.
When the satellite mass is not large enough to open a full gap but creates a partial gap, it is called a propeller structure. 
In this study, we focus on the fully opened gap structures. 

\begin{figure}[h]
\centering
\includegraphics[keepaspectratio, width=0.6\linewidth]{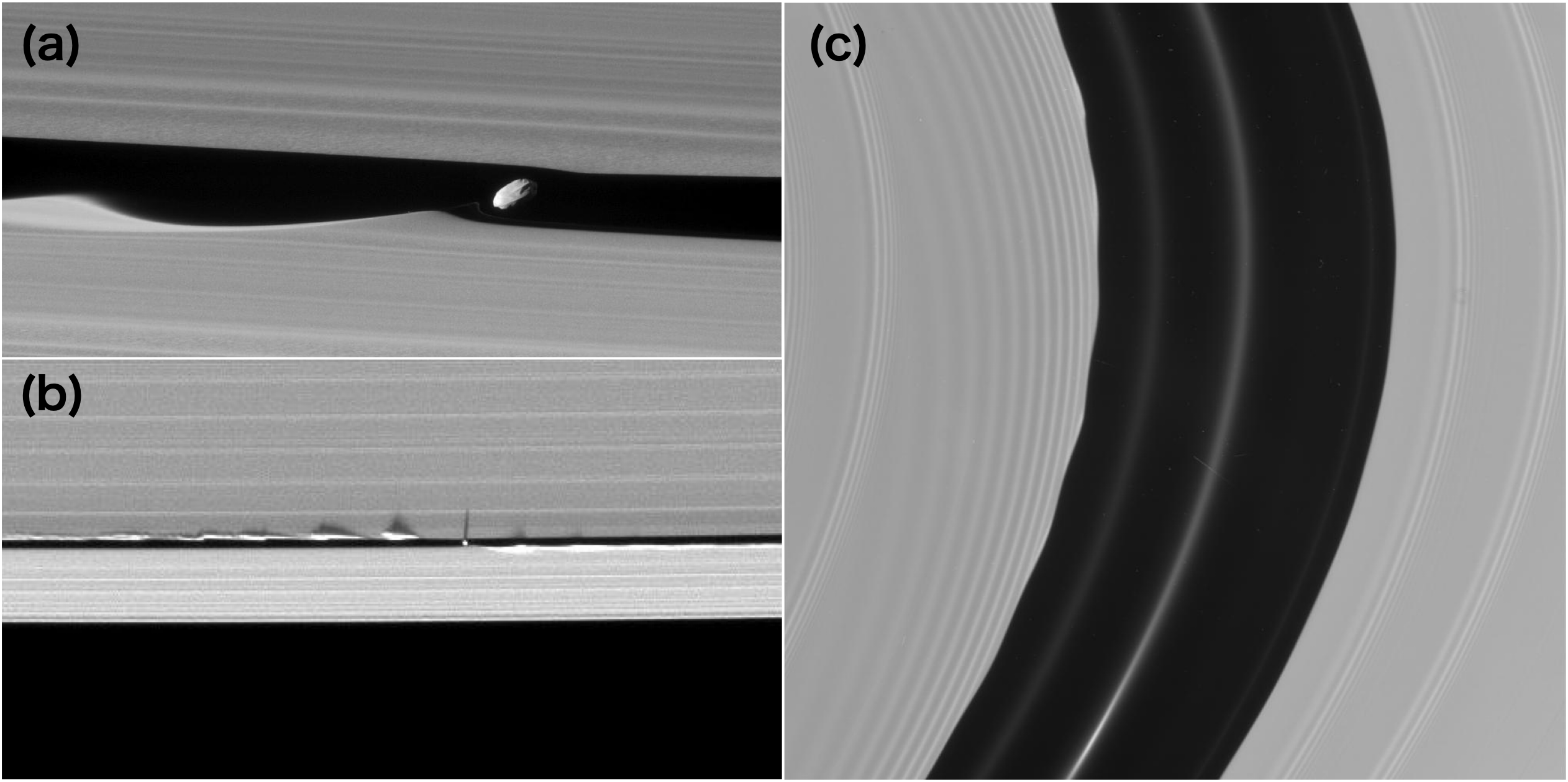}
\caption{(a): A small satellite Daphnis opening the Keeler gap and the density wave induced on the gap edge. 
(b): The vertical structure of density waves induced on the edge of the Keeler gap with a clear shadow cast on the ring. 
(c): The density waves induced at the Encke gap edge by the small satellite, Pan (NASA/JPL/Space Science Institute).}
\label{fig:image}
\end{figure}

Daphnis and Pan are good examples of fully gap-opening satellites (\Figref{fig:image}). The mean physical radius of Daphnis is 3.9$\pm 0.8 \ \rm km$ and it opens the Keeler gap whose inner half-width is 13-20 km and outer half-width is 14-16 km.
The mean physical radius of Pan is 14.2 $\pm$ 1.3 km and it opens the Encke gap with 161 km half-width \citep{Porco:2007, Weiss:2009}. 
The gap width is determined by a balance between the gravitational scattering of the ring particles by the embedded satellite and the viscous diffusion of the ring particles \citep{Lissauer:1981, Petit:1987a, Petit:1987b, Petit:1988}. 

One of the striking features of the gap structures is its sharp edges.
The occultation observations of Saturn's ring revealed the characteristic surface density profile of the gap structure. 
The gap edges created by the satellite are sharply truncated \citep[e.g.,][]{Holberg:1982, Colwell:2009}. 
In other words, the surface density abruptly drops to zero at the gap edge, which is in contrast to the gap structure opened by a planet in a protoplanetary gas disk where the surface density of the gap edge gradually drops while it does not go to zero \citep[e.g.,][] {Kanagawa:2017}. 

\cite{Borderies:1982} introduced the streamline formalism for the description of ring dynamics (for a detailed review of the streamline formalism; see \cite{Longaretti:2016, Longaretti:2017}). 
The formalism is a hybrid of hydrodynamic and celestial mechanics methods by approximating ring particles as fluid based on the Boltzmann equation and utilizing the secular perturbation theory in celestial mechanics for gravitational perturbations to the streamlines from a satellite.  
\cite{Borderies:1982} proposed with the streamline formalism that resonant perturbations by a distant satellite orbiting outside Saturn's A ring, such as Mimas, Janus and Epimetheus, reverse the sign of angular momentum flux locally in the region where the perturbations strongly distort the streamlines of ring particles to propose that it is responsible for the sharp outer edges of the A- and B-rings \citep[e.g.,][]{ElMoutamid:2016, Tajeddine:2017}.  
\cite{Borderies:1989} also proposed the flux reversal caused by a satellite embedded in a gap in the ring such as Pan and Daphnis can be responsible for the sharp edges of the gap. 

More recently, \cite{Gratz:2018, Gratz:2019} solved a radial one-dimensional diffusion equation for a gap profile considering the ring surface density dependence of the ring viscosity. 
They found that when the ring viscosity depends on the surface density ($\Sigma$) with a positive power, the gap edge becomes sharper than when the viscosity is independent of surface density. 
As shown in Section \ref{sec:ring_viscosity}, the ring viscosity is
$\propto \Sigma$ when particle-particle inelastic collision dominates, and 
$\propto \Sigma^2$ when the self-gravity dominates.
It may facilitate the formation of the sharp gap edges.

\cite{Lewis:2000} confirmed that the flux reversal occurs in their local N-body simulation of interactions of a satellite and rings with only inelastic collisions between the ring particles.
Furthermore, \cite{Lewis:2011} revealed through a local N-body simulation that the satellite's perturbations induce ``negative diffusion'' where particles are in high $\Sigma$ regions, which is a counter-intuitive phenomenon. 
They showed that this is due to the damping of particles' eccentricity by frequent mutual inelastic collisions in the density wavefronts where the surface density is extremely high. 

Another striking feature is the vertical mountain-like wall structures associated with the satellite wakes at the gap edge, casting shadows on the ring plane (\Figref{fig:image}b).
A satellite embedded in a gap gravitationally excites steady satellite wakes, particularly at the gap edges (\Figref{fig:image}a). 
The satellite wakes are well understood. 
For example, because the amplitude of the induced epicyclic oscillation depends only on the mass of the satellite and impact parameter \citep{Dermott:1984}, the mass of the satellite can be estimated from the gap width and amplitude \citep[e.g.,][]{Showalter:1986, Weiss:2009}. 
However, it is puzzling that the satellite wakes at the gap edge are accompanied by the vertical mountain-like wall structures.
The height of the vertical structure is estimated to be about $\sim 1$km based on the length of the shadow. 
This is an amazing feature because the typical thickness of Saturn's rings is only about $10$m \citep{Weiss:2009}.

With the streamline model, \cite{Borderies:1985} proposed that particles vertically splash when the streamlines of particles are strongly compressed. 
\cite{Lewis:2000} confirmed the particle splashing in the high-density region with their local N-body simulation. 
On the other hand, \cite{Weiss:2009} developed test-particle models for the satellite wake formation neglecting any particle-particle interactions and concluded that the vertical structure was not caused by the compression of streamlines but by out-of-plane perturbations from the satellite with inclined orbit. 

These features of Saturn's rings identified by the Cassini observations have usually been investigated separately by different theoretical approaches: the streamline model, radial 1D diffusion calculations, and local N-body simulation.
In order to simultaneously and consistently study the gap formation, the sharp gap edges, the satellite wakes, and the vertical walls, we perform high-resolution ($N=1 \times 10^6 - 5 \times 10^6$) global 3D N-body simulation by calculating both gravitational interactions and inelastic collisions among all of ring particles and a satellite.

Pioneering N-body simulations of a planetary ring system with a few hundreds to thousands of particles before the detailed Cassini observations \citep[i.e.,][]{Brahic:1977, Hameen-Anttila:1980, Hanninen:1994, Hanninen:1995} succeeded to derive some basic characteristics of collisional ring systems.
To investigate small-scale ring structures such as wakes caused by the self-gravity of the ring particles, the N-body simulation in a local box with periodic boundary conditions was mostly used \citep[e.g.,][]{Salo:1991, Salo:1992, Salo:1995, Daisaka:1999, Daisaka:2001}. 
After the detailed and striking Cassini observation results were reported, the local simulation was also used \citep[e.g.,][]{Lewis:2000, Lewis:2005, Lewis:2009, Lewis:2011, Michikoshi:2011, Mondino-Llermanos:2023}.
This approach, however, cannot self-consistently address global processes in Saturn's ring such as gap formation and orbital migration of a satellite interacting with the rings \citep[e.g.,][]{Tiscareno:2010}.

Thanks to the recent developments of computational power and N-body simulation codes, it is becoming possible to investigate global phenomena with a global full N-body simulation by calculating gravitational interactions and inelastic collisions among all ring particles and a satellite and with a high enough resolution to reproduce fine structures using enough number of particles. 
\cite{Nakajima:2020} conducted the full N-body simulation with $N\sim 10^5$ to study the orbital expansion of a satellite outside rings that is interacting with the rings with self-gravity wakes to find that the angular momentum transfer by the self-gravity wakes in the rings accelerates the orbital expansion. 
Other global dynamical processes in planetary rings are also important, such as the orbital migration of satellites/moonlets embedded in rings \citep[e.g.,][]{Tiscareno:2010, Crida:2010, Bromley:2013} and the propagation of the density waves due to secular dynamics \citep[e.g.,][]{Hahn:2007, Hahn:2008}. 

As a first attempt of our series of papers, we conduct the global full N-body simulation with $N=10^6-10^7$ to simultaneously and consistently calculate the gap formation, the sharp gap edges, the satellite wakes, and the vertical walls that have mostly been separately simulated by different approaches. 
Even with $N=10^6-10^7$, we need to use satellite and ring particles much larger than those in the real Saturn's satellite-ring system.
As we show in the following sections, the mechanisms we focus on here are scaled with the Hill radius of the satellite. 
At least for these points, the results of our global simulation can be applied to the real system. 
We also discuss that the relative importance of collisional and self-gravitational viscosities of the ring particles in our simulation and the real system (\Secref{Comparison}).

In \Secref{Method}, we describe the simulation method, settings and some important parameters. 
\Secref{Result} presents our main results. 
The satellite wakes, the sharp gap edges, and the vertical walls are simultaneously reproduced by a single simulation. 
We show a detailed analysis of the formation of these structures.   
In \Secref{Discussion}, we discuss the justification of our global simulation by comparing it with the real system for the subjects addressed in this paper.
\Secref{Conclusion} consists of a summary of this paper and some future perspectives.

\section{Method and Simulation Settings} \label{Method}

\subsection{Self-gravity wakes}

To explain our simulation settings and parameters, and to help later analysis, we summarize the results of local N-body simulations on the self-gravity wakes \citep{Salo:1992, Salo:1995, Daisaka:1999,Daisaka:2001} here and ring viscosity in Section \ref{sec:ring_viscosity}.

The self-gravitational instability of a disk is evaluated with the Toomre's parameter $Q$ \citep{Toomre:1964}:
\begin{equation}
Q=\frac{\Omega\sigma_r}{3.36G\Sigma},
\label{eq:Q}
\end{equation}
where $\sigma_r$ is the radial velocity dispersion of particles, $G$ is the gravitational constant, $\Omega$ is the local Keplerian angular velocity, and $\Sigma$ is the disk surface mass density. If $Q\lesssim2$ is satisfied, self-gravity wakes appear in the disk, and the ring's viscosity is enhanced. The local N-body simulations showed
\begin{equation}
\sigma_r\sim
\left\{
\begin{array}{ll}
2 R \Omega & [r_{\rm h}^*\lesssim0.5] \\
\displaystyle \sqrt{Gm/R} &  [r_{\rm h}^*\gtrsim0.5]
\end{array}
\right.,
\label{eq:sigma}
\end{equation}
where $R$ and $m$ are the physical radius and the mass of particles, respectively, and
\begin{align}
r_{\rm h}^* = \frac{r_{\rm H}}{2 R} \simeq 1.07 \frac{r}{r_{\rm R}}.
\end{align}
where $r_{\rm H}$ and $r_{\rm R}$ are the Hill and the Roche limit radii defined by
\begin{align}
r_{\rm H} & = \left(\frac{2 m}{3M_{\rm p}}\right)^{1/3} r,
\label{eq:rH} \\
r_{\rm R} & = 2.456 \left(\frac{\rho_{\rm p}}{\rho} \right)^{1/3} R_{\rm p}, \label{eq:rR}
\end{align}
where $\rho_{\rm p}$ and $\rho$ are bulk densities of the planet and the ring particles, $R_{\rm p}$ is the physical radius of the planet, and $M_{\rm p}$ is the mass of the planet.
While in inner ring regions $r \ll r_{\rm R}$ ($r_{\rm H} \ll R$), the self-gravity is not effective, the self-gravity wakes develop in outer ring regions ($r \sim r_{\rm R}$) in the case of relatively high optical depth.
Substituting \Eqref{eq:sigma} into \Eqref{eq:Q}, the condition for the appearance of self-gravity wake $Q\lesssim2$ is 
\begin{equation}
\tau\gtrsim
\left\{
\begin{array}{ll}
0.08r_{\rm h}^{*-3} &  [r_{\rm h}^*\lesssim0.5]\\
0.2r_{\rm h}^{*-3/2} & [r_{\rm h}^*\gtrsim0.5]
\end{array}
\label{eq:selfgrav}
\right.,
\end{equation}
where $\tau$ is the geometrical optical depth of rings, which is related to the surface density as $\tau=\pi R^2\Sigma/m$. 
If the satellite is set at $r = 0.66 r_{\rm R}$, $r_{\rm h}^*\simeq0.7$ and the condition is $\tau\gtrsim0.34$.

\subsection{Ring viscosity} \label{sec:ring_viscosity}

Next, we summarize viscosity (angular momentum transfer rate) in a ring system. 
In a ring system, the viscosity is divided into the angular momentum transfer by inelastic collisions of the ring particles ($\nu_{\rm coll}$), gravitational torque from wakes caused by the ring particle self-gravity ($\nu_{\rm grav}$), and the associated translation carried by the particle motions ($\nu_{\rm trans}$). 
The ring viscosity has been investigated both by analytical approaches and N-body simulations \citep[e.g.,][]{Goldreich:1978, Wisdom:1988, Salo:1991, Daisaka:2001, Morishima:2006}. 

The expression for the collisional and translational viscosity without the effect of the self-gravity was analytically derived as \citep{Goldreich:1978,Araki:1986} :
\begin{equation}
\nu_{\rm coll}\simeq R^2\Omega\tau,
\label{eq:coll}
\end{equation}
\begin{equation}
\nu_{\rm trans}=\frac{\sigma_{r}^2}{2\Omega}\qty(\frac{0.46\tau}{1+\tau^2}).
\label{eq:trans}
\end{equation}

The viscosity considering the effect of the self-gravity was investigated by N-body simulations \citep[e.g.,][]{Daisaka:2001, Takeda:2001}. 
In the case of the optically thin ring ($\tau\ll1)$, the total viscosity $\nu_{\rm tot}$ is derived as:
\begin{equation}
    \nu_{\rm tot}=2.75Ir_{\rm h}^*\tau R^2\Omega,
    \label{eq:viscosity_thin}
\end{equation}
where $I$ is a non-dimensional factor. 
\cite{Petit:1987a} obtained $I=5.78$ for $r_{\rm h}^*=0.70$ and $\varepsilon=0.1$, where $\varepsilon$ is the restitution coefficient.
Using the local N-body simulation, \cite{Daisaka:2001} found that, when the self-gravity wakes develop, $\nu_{\rm trans}$ is enhanced by the collective motion of the self-gravity wakes. 
They showed that $\nu_{\rm trans}\simeq \nu_{\rm grav}\gtrsim\nu_{\rm coll}$ and the total viscosity ($\nu_{\rm tot}$) is given by
\begin{equation}
\nu_{\rm tot}\simeq\nu_{\rm grav} + \nu_{\rm trans} \simeq C \frac{G^2\Sigma^2}{\Omega^3},
\label{eq:grav}
\end{equation}
where $C = 26 \,(r_{\rm h}^*)^5 = 26 \,(r_{\rm H}/2 R)^{5} = 26 \,(r/0.93 r_{\rm R})^5$ represents the radial dependence in the rings.

\subsection{N-body simulation}

We use an open-source N-body simulation code \texttt{n-body-with-center}, which was developed for a simulation of planetary ring system \footnote{
\texttt{n-body-with-center} is available on the GitHub page of J.Makino (https://github.com/jmakino/nbody-with-center).
} \citep{Iwasawa:2020}. 
In this code, the soft-sphere model \citep{Salo:1995, Mondino-Llermanos:2022} is adopted for inelastic collisions of particles. 
Inelastic collisions are represented by adding a restoring harmonic force and an energy dissipation force during collisions. 
The collisional force exerted to a particle is expressed as
\begin{equation}
    \vb{F}_{\rm coll}(\alpha)=
    \left\{
    \begin{array}{ll}
    (k_1d+k_2\dot{d})\vb{e} & [d>0 \ : \ {\rm{during \ a \ collision}}]\\
    0 & [d\leq 0 \ : \ {\rm{others}}]
    \end{array}
    \right.
    ,
\end{equation}
where $d$ is the penetration depth during the collision and $\vb{e}$ is the unit vector in the direction joining the particle centers. 
The first term is the repulsive force to reverse the normal component of the relative velocity and the second term is the energy dissipation due to the inelastic collision. 
The spring and dissipation constants $k_1$ and $k_2$ are related to the duration of the collision $T_{\rm imp}$ and the restitution coefficient $\varepsilon$:
\begin{align}
    \frac{k_2}{\mu}&=-\frac{2}{T_{\rm imp}}\log{\varepsilon}, \\
    \frac{k_1}{\mu}&=\frac{\pi^2}{T_{\rm imp}^2}+\frac{1}{4}\qty(\frac{k_2}{\mu})^2,
\end{align}
where $\mu$ is the reduced mass of the collision pair and $T_{\rm imp}$ is the duration time of the collision (half of the oscillation period) \citep{Dilley:1993}. 
In this study, we set $T_{\rm imp}=T_{\rm K}/(2^4\cdot2\pi)\sim10^{-2}T_{\rm K}$ and $\varepsilon=0.1$, where $T_{\rm K}$ is the Kepler time at the outer edge of the ring. 
We integrate the equation of motion with the leapfrog method and the time step we use is $dt=T_{\rm K}/(2^{10}\cdot2\pi)\sim1.6\times10^{-4}T_{\rm K}$. 
This code uses Framework for Developing Particle Simulator (FDPS) \citep{Iwasawa:2016,Namekata:2018}. 
The gravitational interaction of particles is calculated with the Barnes-Hut tree scheme \citep{Barnes:1986} available in FDPS. 
Saturn’s $J_2$ potential causes a secular non-Keplerian precession on the particle’s eccentric orbit, and its typical timescale is much longer than that of the mechanism that we are concerned with here (see \Secref{Result}).
Thus, we do not consider the $J_2$ potential.

In order to split the gap formation mechanism and the time evolution of the orbital elements of the satellite, we fix the orbit of the satellite in Keplerian circular orbit whose orbital radius is $0.66r_{\rm R}$, where $r_{\rm R}$ is the Roche limit radius of Saturn 
with $\rho=0.5 \ \rm{g/cm^3}$ and $\rho_{\rm p} = 0.7 \ \rm{g/cm^3}$
(Eq.~\ref{eq:rR}). 
We will study the orbital evolution of the satellite by the interaction with the disk in the next paper (see also \Secref{Conclusion}).

In this study, we assume that all particles are spherical and have the equal size. 
Each particle can be considered to be a super-particle that represents a swarm of ring particles, which could be justified because the collective motion due to, for example, self-gravity wakes, is dominant in Saturn's rings \citep{Colwell:2006}. 
We set the bulk density of a particle $0.5 \ \rm{g/cm^3}$ and determine its radius from its bulk density and mass. 

The number of particles we use is $N=1\times10^6-5\times10^6$.
To prevent particles from accumulating outside $r_{\rm R}$, we place these particles in a radial range of $0.41 \, r_{\rm R}$ to $0.82 \, r_{\rm R}$ with a constant surface density. 
The initial orbital eccentricity and inclination of these particles are given to follow a Rayleigh distribution with the root-mean-square value of $\langle e^2\rangle^{1/2}=\langle i^2\rangle^{1/2}=0.05$. 
Saturn's surface corresponds to $~0.36\, r_{\rm R}$. 
If a particle reaches the surface of Saturn, it is removed from the simulation.

\Tabref{tab:run_summary} shows the parameter sets, the number of particles $N$, inidividual particle mass $m$ (equal mass), initial geometrical optical depth $\tau_0$, the total ring mass $M_{\rm ring}$ and satellite mass $M_{\rm s}$ in each model. 
In the following section, we use the normalized quantities with tilde, $\tilde{t}=t/\Omega^{-1}= t/(T_{\rm K}/2\pi)$, $\tilde{r}=r/r_{\rm R}$, $\tilde{m} = m/M_{\rm p}$, $\tilde{M}_{\rm ring} = M_{\rm ring}/M_{\rm p}$, and $\tilde{M}_{\rm s} = M_{\rm s}/M_{\rm p}$.

\begin{table*}[h]
 \caption{Summary of our models : $N$ is the number of particles, $\tilde{m}$ is the mass of a particle normalized by Saturn's mass, $\tau_0$ is the initial optical depth, $\tilde{M}_{\rm ring}$ is the normalized total ring mass and $\tilde{M}_{\rm s}$ is the normalized mass of the satellite}
 \centering
  \begin{tabular}{ccccccc}
   \hline
      & $N$ & $\tilde{m}$ & $\tau_0$ & $\tilde{M}_{\rm ring}$ & $\tilde{M}_{\rm s}$ \\
   \hline \hline
   Model 1 & $3\times10^6$ & $4\times10^{-11}$  & 0.11 & $1.2 \times 10^{-4}$ & $5\times10^{-6}$\\
   Model 2 &     ...       &       ...          &  ... & ... & $1\times10^{-6}$ \\
   Model 3 & $1\times10^6$ & $2\times10^{-10}$  & 0.10 & $2 \times 10^{-4}$ & $1.5\times10^{-5}$\\
   Model 4 &     ...       &       ...          &  ... & ... & $1\times10^{-5}$ \\
   Model 5 &     ...       &       ...          &  ... & ... & $8\times10^{-6}$ \\
   Model 6 &     ...       &       ...          &  ... & ... & $5\times10^{-6}$ \\
   Model 7 &     ...       &       ...          &  ... & ... & $3\times10^{-6}$ \\
   Model 8 &     ...       &       ...          &  ... & ... & $1\times10^{-6}$ \\
   Model 9 &     ...       &       ...          &  ... & ... & $8\times10^{-7}$ \\
   Model 10 & $3\times10^6$ & $2\times10^{-10}$  & 0.31 & $6 \times 10^{-4}$ & $1\times10^{-5}$ \\
   Model 11& $5\times10^6$ & $2\times10^{-10}$  & 0.52 & $1 \times 10^{-3}$ & $2\times10^{-5}$ \\
   \hline
   \label{tab:run_summary}
  \end{tabular}
\end{table*}

\section{Results}\label{Result}

\subsection{Satellite wakes}\label{DensityWave}

\begin{figure}[h]
\centering
\includegraphics[width=0.6\linewidth]{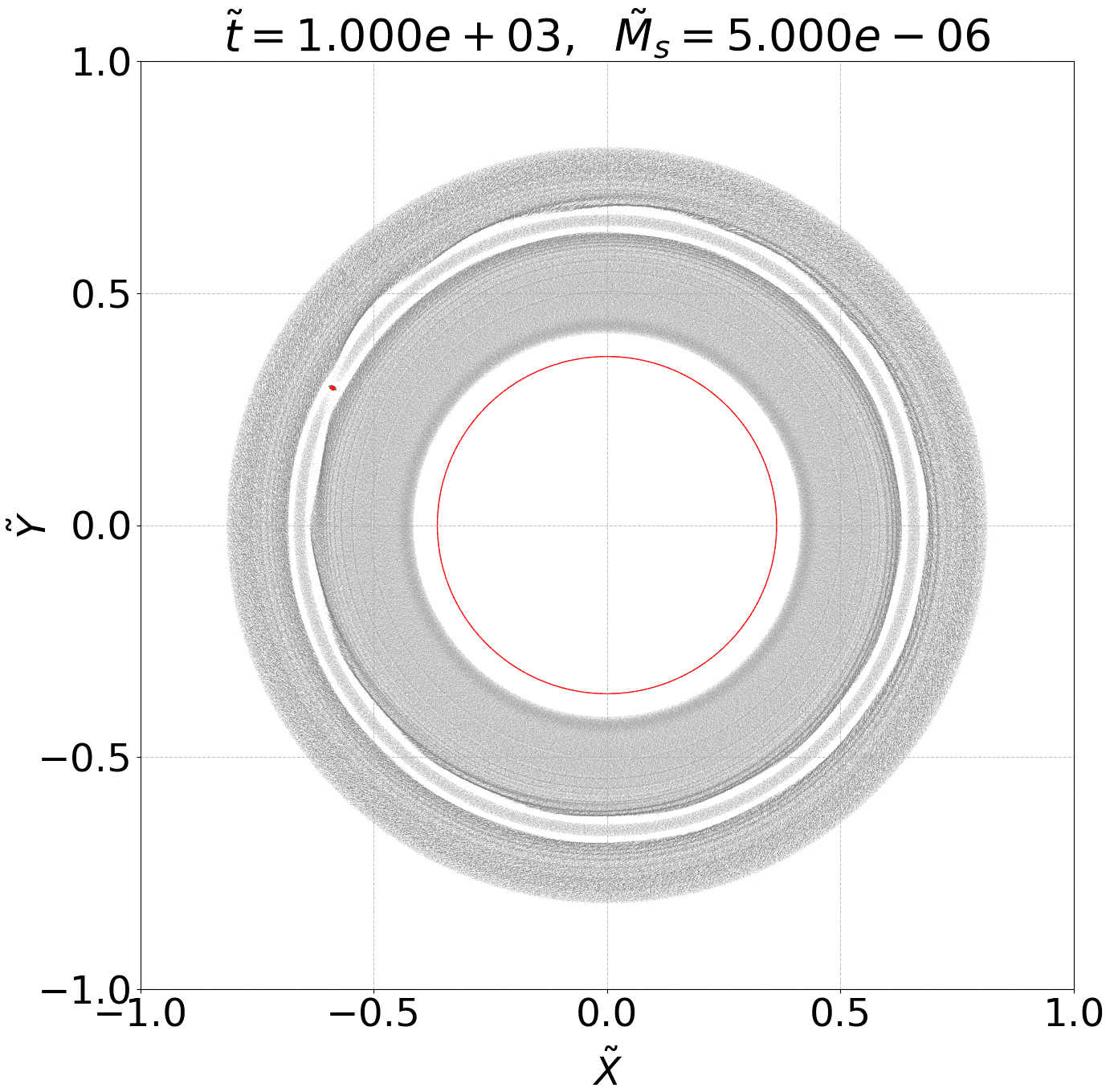}
\caption{The snapshot of the ring at $\tilde{t}=1000$ (Model 1). The circle marked by the red curve is the surface of Saturn. The red dot shows the satellite position.}
\label{fig:snap}
\end{figure}

\begin{figure*}[h]
\centering
\includegraphics[width=0.8\linewidth]{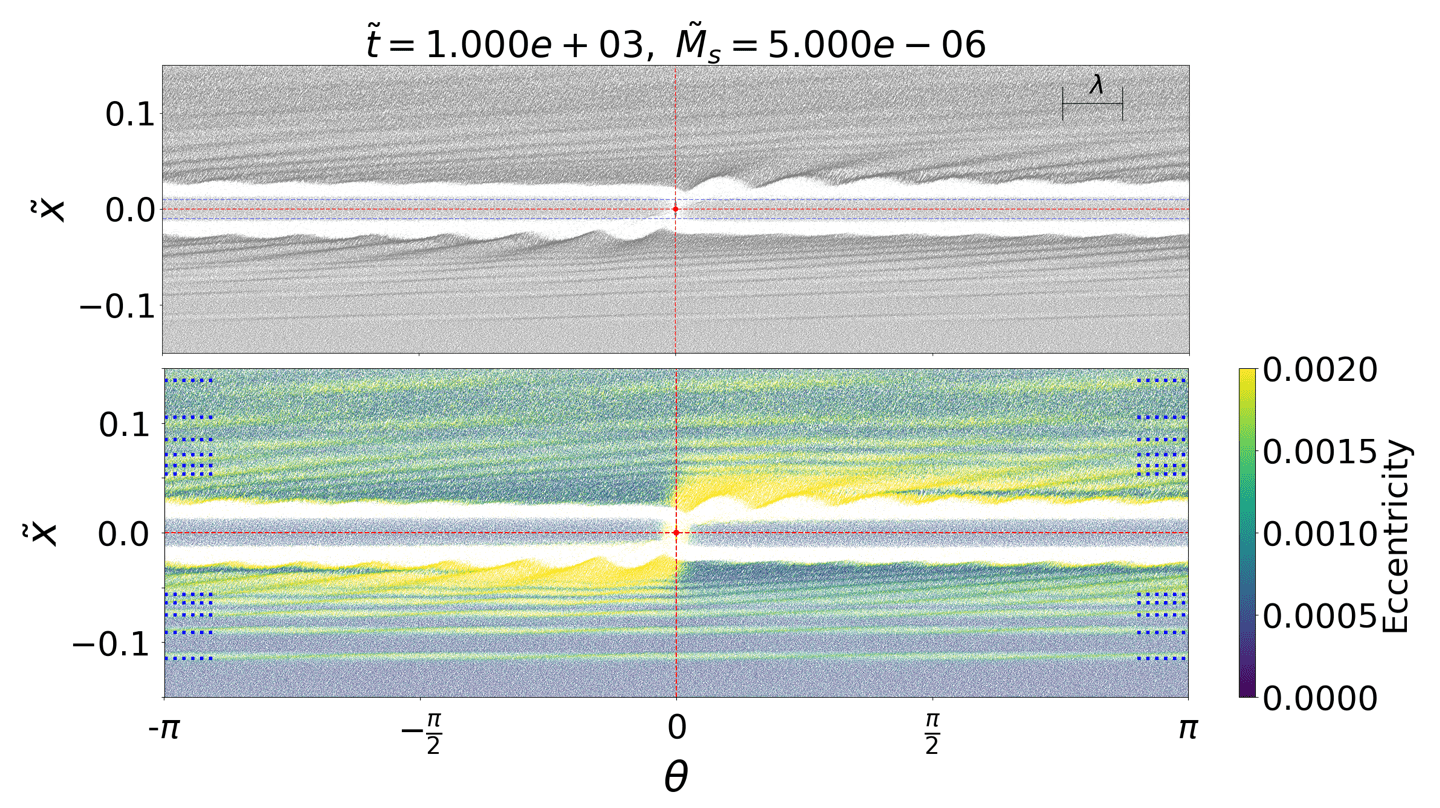}
\caption{The snapshots of the satellite wakes excited near the gap edge (Model 1). The satellite is at the origin. The blue dotted lines in the upper panel represent the estimated width of the horseshoe region \citep{Dermott:1980}. The bar in the upper right of the upper panel is the theoretical azimuthal wavelength of the satellite wake \citep{Seiss:2010}. 
The color of the dots in the lower panel represents the eccentricity of each particle. The blue dotted lines in the left and right sides in the lower panel represent the positions of the first-order Lindblad resonances, $r_{\rm L}=(1+(1/m))a_{\rm s}$ and $m = -4, -5, -6, -7, -8, 8, 7, 6, 5, 4, 3$ from inner to outer lines, respectively.}
\label{fig:wake}
\end{figure*}

Figure~\ref{fig:snap} is a snapshot of the ring particles projected on the midplane at $\tilde{t}=1000$ of Model 1 ($\tilde{M}_{\rm s}=5\times10^{-6}$). 
The satellite, which is shown as a red dot, opens a wide gap and excites a tightly-wound density wave pattern which is similar to that was observed near the edge of the Encke gap by Cassini (\Figref{fig:image}). 
Figure~\ref{fig:wake} is snapshots of the same model stretched along the azimuthal direction from the global coordinates to the rectangular coordinates. 
The vertical and horizontal axes represent radial distances $\tilde{x}$ and azimuthal angle $\theta$ relative to the satellite. 
The satellite at the origin is shown as a red dot. 
In the upper (lower) area particles flow from left to right (right to left) by the Keplerian shear motions. 
Wavy patterns at the gap edge are excited by the gravitational perturbation from the satellite, referred to as ``satellite wake'' hereafter. It should be noted that the satellite wakes survive until the next encounter with the satellite without completely damping out because of small synodic period in our global simulation. 
In contrast, in the case of Encke gap, \cite{Showalter:1986} pointed out that the satellite wakes only survive for about $270^{\circ}$ of azimuth from the position of Pan.

\begin{figure}[h]
\centering
\includegraphics[width=0.5\linewidth]{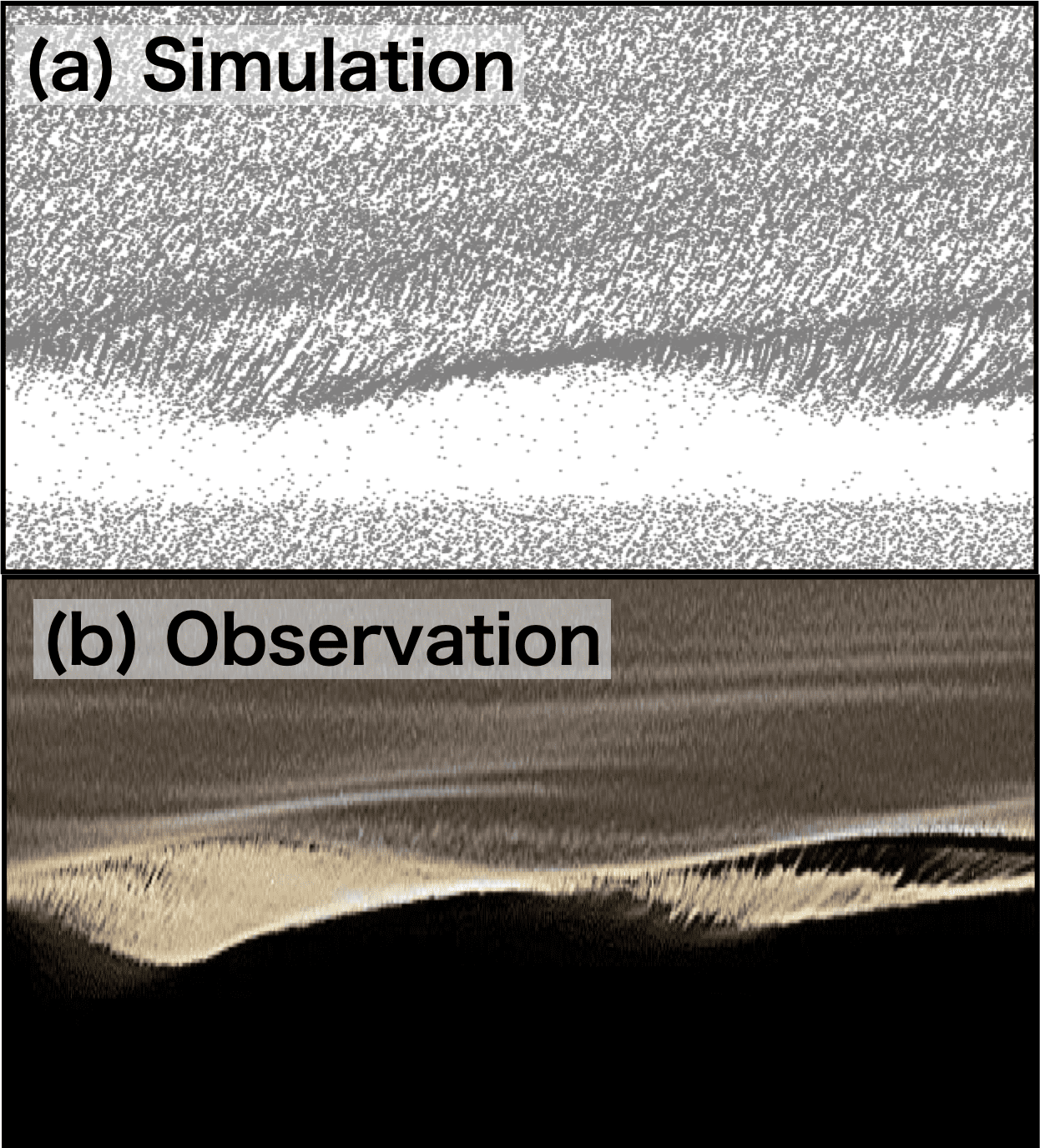}
\caption{The comparison between (a) N-body simulation and (b) Cassini's Grande Finale Observation \citep{Tiscareno:2019}. The observed superposition of satellite wakes and self-gravity wakes is demonstrated in our N-body simulation.}
\label{fig:wake_zoom_comp}
\end{figure}

The azimuthal wavelength of the satellite wake is $\tilde{\lambda}=3\pi \tilde{x}$ \citep[e.g.,][]{Seiss:2010}. 
Its corresponding length is shown in the upper right of the upper panel of \Figref{fig:wake}. 
The wavelength of the created satellite wakes in the simulation is consistent with the estimated length. 
Downstream of the satellite, ropy structures created by self-gravitational instability can be seen in the wave crests, which are not clear in the first wave crest. 
These features were observed during the close-range flybys of the main ring in the Cassini's Grand Finale \citep[][see \Figref{fig:wake_zoom_comp} for the comparison]{Tiscareno:2019}.
In the Grand Finale observation, macroscopic clumps of ring particles and a gap between the clump and the wave crest in the third wave crest were found. 
However, they are not found in our simulation, thus they could be caused by the deviation from circular and coplanar orbit (i.e. the vertical or epicycle motion) of the satellite as inferred in \cite{Tiscareno:2019}, which is left for the future works.

Another component of wakes originating from the self-gravity of particles appears when the disk is gravitationally unstable (see \Figref{fig:wake_zoom_comp} and  \Figref{fig:wake_highTau_zoom}).
We refer to this component as ``self-gravity wake'' to distinguish it from the satellite wake.
We will show the results where the self-gravity wake is strongly excited in \Secref{sec:StrongGravityWake}. 

\begin{figure}[h]
\centering
\includegraphics[width=0.6\linewidth]{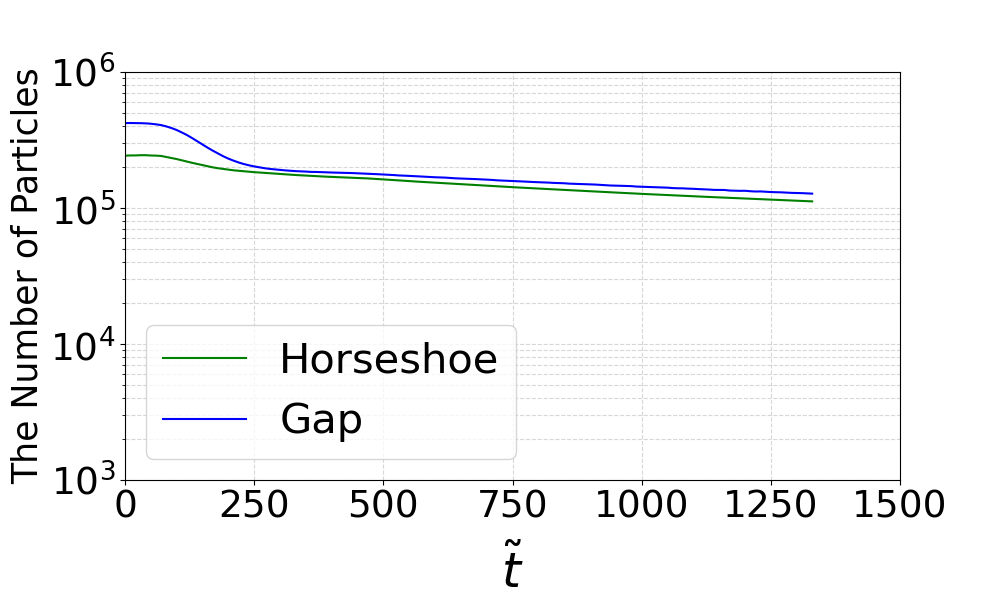}
\caption{Time evolution of the number of particles in the gap and horseshoe region in Model 1. We define the gap and the horseshoe width based on \Eqref{eq:gap_width} and \Eqref{eq:horseshoe}, respectively, and count the number of particles in these regions.}
\label{fig:horseshoe}
\end{figure}

Figures~\ref{fig:snap} and \ref{fig:wake} show that many particles remain in the horseshoe region. 
The radial half-width of the horseshoe orbit band $W_{\rm h}$ is \cite[e.g.,][]{Dermott:1981}:
\begin{equation}
    W_{\rm h}\simeq 1.3 \tilde{M}_{\rm s}^{1/3}a_{\rm s} \sim \, 2r_{\rm H,s},
    \label{eq:horseshoe}
\end{equation}
where $a_{\rm s}$ is the semi-major axis of the satellite
and $r_{\rm H,s}=(M_{\rm s}/3 M_{\rm p})^{1/3}a_{\rm s}$
is its Hill radius. 
The predicted horseshoe band is plotted by blue dotted lines in the upper panel in \Figref{fig:wake}.
Figure~\ref{fig:horseshoe} shows the time evolution of the number of particles remaining in the gap (blue curve) and in the horseshoe region (red curve) in Model 1. We define the gap width and the horseshoe width based on \Eqref{eq:horseshoe} and \Eqref{eq:gap_width} (see \Secref{GapWidth}), respectively, and counts the particles in these regions. In Model 1, particles are radially uniformly distributed at $\tilde{t}=0$. 
Initially, $\sim3\times10^5$ particles were in the horseshoe orbits. 
The particles gradually diffused out, but $\sim1\times10^5$ particles still remain at $\tilde{t}=1000$ (\Figref{fig:horseshoe}). As \Figref{fig:image}c shows, the observation shows that many ring particles remain in the horseshoe orbit band along Pan.
These particles could also provide materials to form ringlets observed in the Encke gap \citep{Showalter:1991, Hedman:2012}. 

Figure~\ref{fig:horseshoe} also shows that typical timescale of gap opening is $\tilde{t}_{\rm gap}\sim200$. On the other hand, the synodic period of the satellite and a particle in the gap edge is $\tilde{t}_{{\rm{syn}}}=2\pi/|\tilde{\Omega}_s - \tilde{\Omega}| \sim 76$. 
Thus, several scatterings open the gap and the gap width should be scaled only by the Hill radius of the satellite (see \Secref{GapWidth} in a more detailed discussion).

In the lower panel of \Figref{fig:wake}, each particle's eccentricity is shown as the color of the dots. 
The blue dotted lines in the left and right sides represent the first-order Lindblad resonances with the satellite, $r_{\rm L}=(1+(1/m))a_{\rm s}$, where $m = -4, -5, -6, -7, -8, 8, 7, 6, 5, 4, 3$ from inner to outer lines, respectively.
The eccentricities of the particles that undergo relatively close encounters are quickly excited at encounters with the satellite (blue color dots are turned to yellow), but they are gradually damped by the particle collisions after the encounters, compared with the excitation timescale. 
Even for distant encountering particles, the eccentricities are relatively highly excited at the Lindblad resonances (along the blue dotted lines). 

\begin{figure}[h]
\centering
\includegraphics[width=\linewidth]{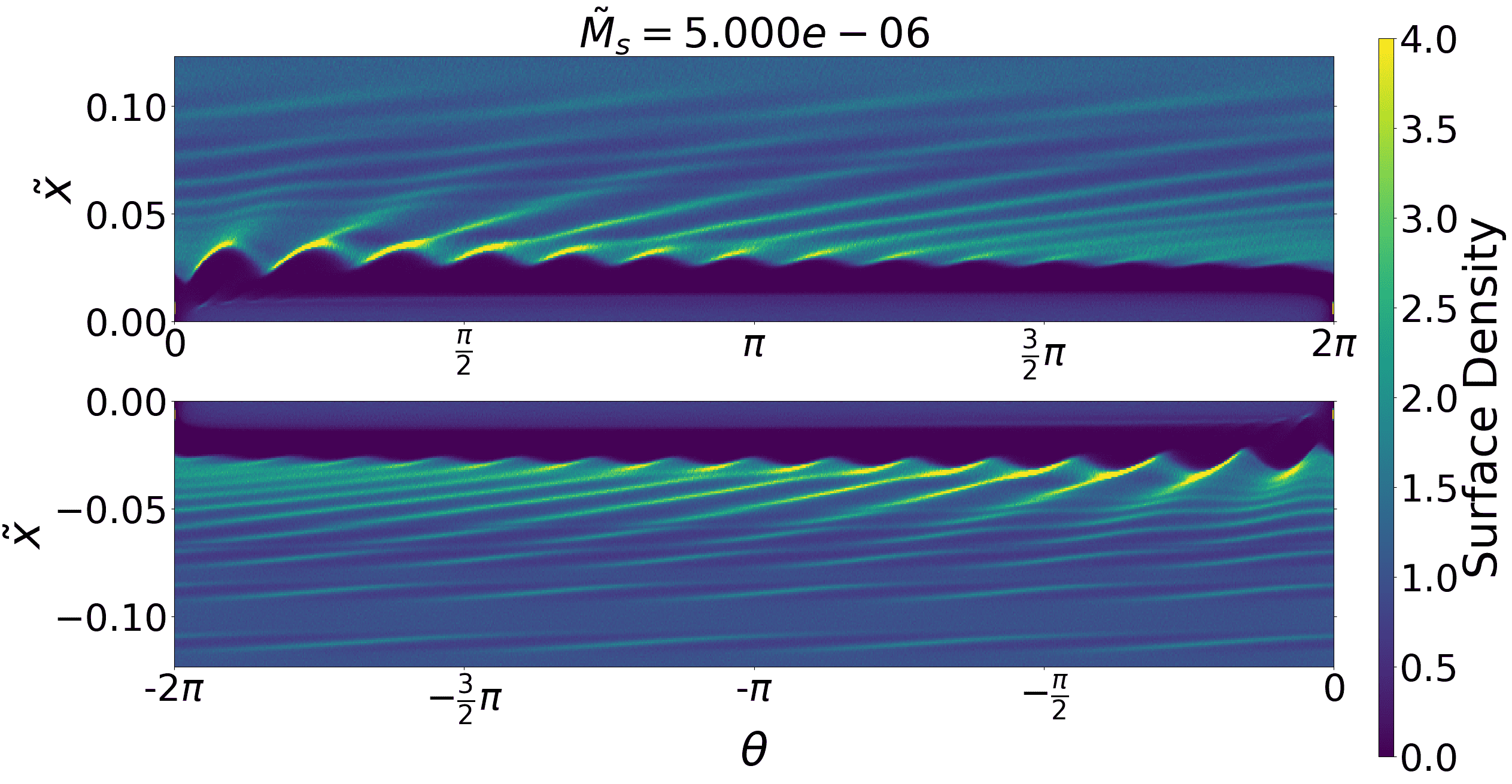}
\caption{The steady surface density map of ring particles in Model 1
after the distribution is sufficiently relaxed $\tilde{t}=1000-1052$.
The surface density is scaled by the initial value and calculated by Eq.~(\ref{eq:Sigma_map}) with the time interval for the integral $\Delta \tilde{t}=1$. The spatial grids are $\Delta \theta=2\pi/900$ in azimuth and $\Delta \tilde{x} =0.4/500$ in radius.
In the upper and lower panels, the satellite is at the bottom left corner
and at the top right corner, respectively.  
}
\label{fig:heatmap}
\end{figure}

\begin{figure*}[h]
\centering
\includegraphics[width=0.9\linewidth]{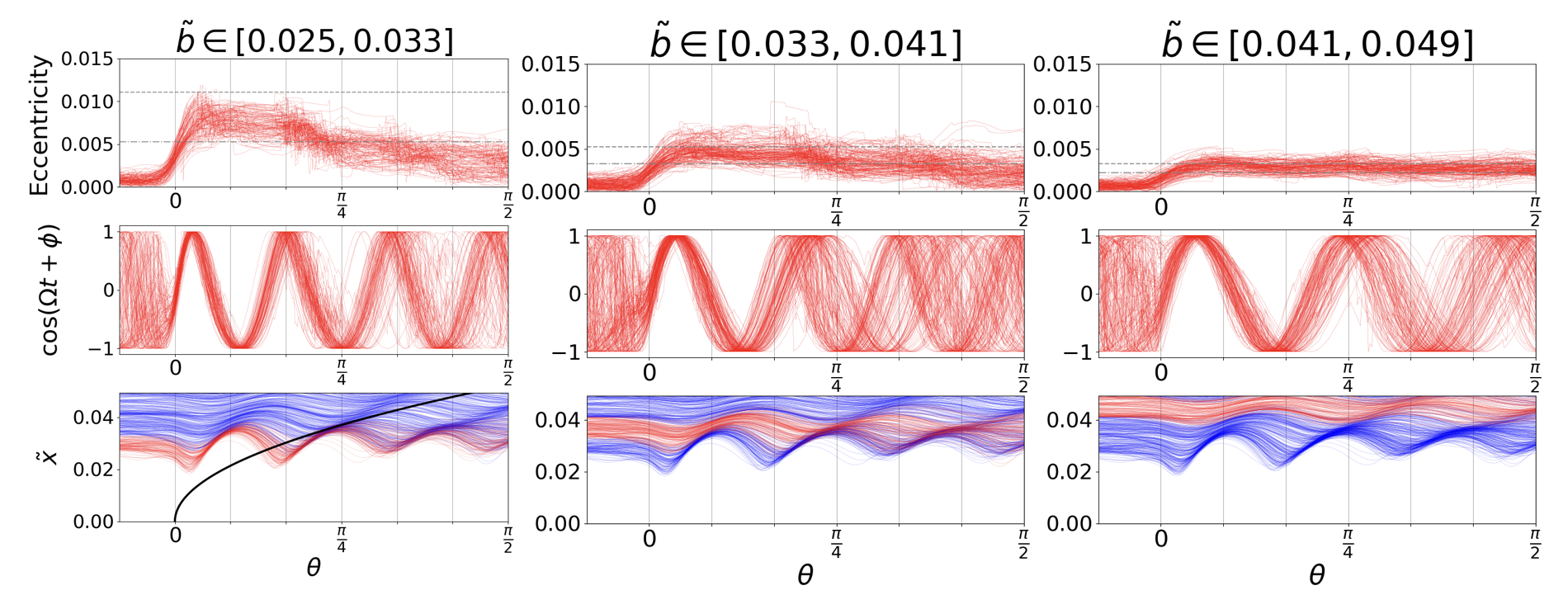}
\caption{Eccentricity, phase of epicyclic oscillation, and streamlines of particles with impact parameters $\tilde{b}\in[0.025, 0.033], [0.033, 0.041], [0.041, 0.049]$ are shown from left to right, respectively. 
The upper panels: Eccentricity induced by gravitational scattering by the satellite. The dotted and dashed lines are the upper and lower estimated values of the excited eccentricity during one encounter with the satellite, which are calculated by \Eqref{eq:weiss}. The middle panels: Epicycle oscillation of each particle $\cos{(\Omega t+\phi)}$. The oscillation phases are synchronized due to the gravitational scattering by the satellite. The lower panels: Particle trajectories. The red-colored trajectories are used for the upper and middle plots. The black curve in the left lower panel shows the location of streamlines crossing by \Eqref{eq:theta_crit} \citep{Showalter:1986}.}
\label{fig:streamline}
\end{figure*}

These features are more clearly shown in the surface density map in \Figref{fig:heatmap}. 
This figure shows the normalized surface density at each grid time-averaged over the synodic period between the satellite and particles near the gap edge ($\tilde{x} =0.05$) that is close to the gap half width after the surface density distribution is relaxed,
\begin{align}
\left\langle\frac{\Sigma(r, \theta)}{\Sigma_0}\right\rangle = \frac{1}{t_1 - t_0}\int_{t_0}^{t_1}\frac{\Sigma(r, \theta)}{\Sigma_0}dt,
\label{eq:Sigma_map}
\end{align}
where $(r,\theta)$ is the rotating cylindrical coordinate system $(r=a_{\rm s} + x)$ and $\Sigma_0$ is the initial surface density. 
The surface density has peaks at each satellite wake edge. 
In these regions, inelastic collisions occur frequently, and then the amplitude of the wakes is damped as the particles move apart from the satellite (see also \Figref{fig:streamline}). 
Density spiral arms also emerge connecting satellite wakes.
The Kepler shear tightly winds these arms. 

To investigate the orbital evolution of individual particle motions in the outer edge fronts, we extracted the trajectories of particles after the encounters. 
Figure~\ref{fig:streamline} shows the eccentricity, epicycle oscillation $\cos{(\Omega t+\phi)}$ and their trajectories, where $\phi$ is the epicycle phase constant. 
The left, middle, and right panels show the trajectories with impact parameters normalized with the Roche limit radius of Saturn $\tilde{b}$ $(= \tilde{x}$) in the ranges of $\tilde{b}\in[0.025, 0.033], [0.033, 0.041],$ and $ [0.041, 0.049]$, respectively. 
The trajectories in the individual ranges are highlighted by the red curves in the bottom panels. 
Theoretical estimate of excited eccentricity by a single encounter with a satellite was derived by \cite{Weiss:2009}:
\begin{equation}
e_{\rm H}=\frac{6.72}{b_{\rm H}^2} + \frac{4}{b_{\rm H}^3} + \frac{1.1\times10^6}{b_{\rm H}^{14}},
\label{eq:weiss}
\end{equation}
where $e_{\rm H} = e/(r_{\rm H,s}/a_{\rm s})$, $b_{\rm H} = b/r_{\rm H,s}$, $b$ is the impact parameter.
On the upper panel of \Figref{fig:streamline}, upper values (corresponding to the lower boundary of $\tilde{b}$) and lower values (corresponding to the upper boundary of $\tilde{b}$) of excited eccentricity estimated by \Eqref{eq:weiss} are also shown as dashed and dotted lines. 

The upper panel shows that the particles' eccentricities rapidly increase due to the gravitational perturbations by the satellite. 
The excited eccentricity is roughly consistent with the estimate of \Eqref{eq:weiss}; some particles obtain higher eccentricity because these particle's eccentricity before the encounter is not negligible, which is different from the assumption in \Eqref{eq:weiss}.
After that, inelastic collisions between particles occur frequently in the high-density region where the streamlines are strongly compressed (see also \Figref{fig:heatmap}), and the eccentricities and epicyclic motions gradually decay.
The location of streamline crossing, referred to as $\theta_{\rm crit}$, is derived as a function of particle's forced eccentricity and its distance from the satellite by \cite{Showalter:1986}:
\begin{equation}
    \theta_{\rm crit}=\frac{3\delta^2}{2e},
    \label{eq:theta_crit}
\end{equation}
where $\delta=\tilde{x}/\tilde{a}_{\rm s}$. 
The streamline crossing location by \Eqref{eq:theta_crit} with $e=0.006$ is shown as a black curve in \Figref{fig:streamline}.
The streamlines cross in the second wake and inelastic collisions are triggered there, thus \Eqref{eq:theta_crit} is consistent with our results.
Even inelastic collisions usually result in diffusion as long as $e$ is relatively small. 
However, the mutual inelastic collisions coupled with the excitation by the satellite causes counter-intuitive ``negative diffusion''.
This was also suggested by the previous local N-body simulation mainly in the context of the formation of compact ringlets \citep{Lewis:2011}.
Our simulation shows that this mechanism also works at gap edges. 
The middle panel shows that the epicyclic oscillation phase of particles is synchronized just after the gravitational scattering by the satellite \citep{Yoshida:2021, Yoshida:2023}, and the synchronization is gradually broken by the Kepler shear, which leads to more frequent collisions and creates the vertical wall structure (see \Secref{Vertical}).  

\subsection{Sharp edge}

\begin{figure*}[h]
\centering
\includegraphics[keepaspectratio, width=0.9\linewidth]{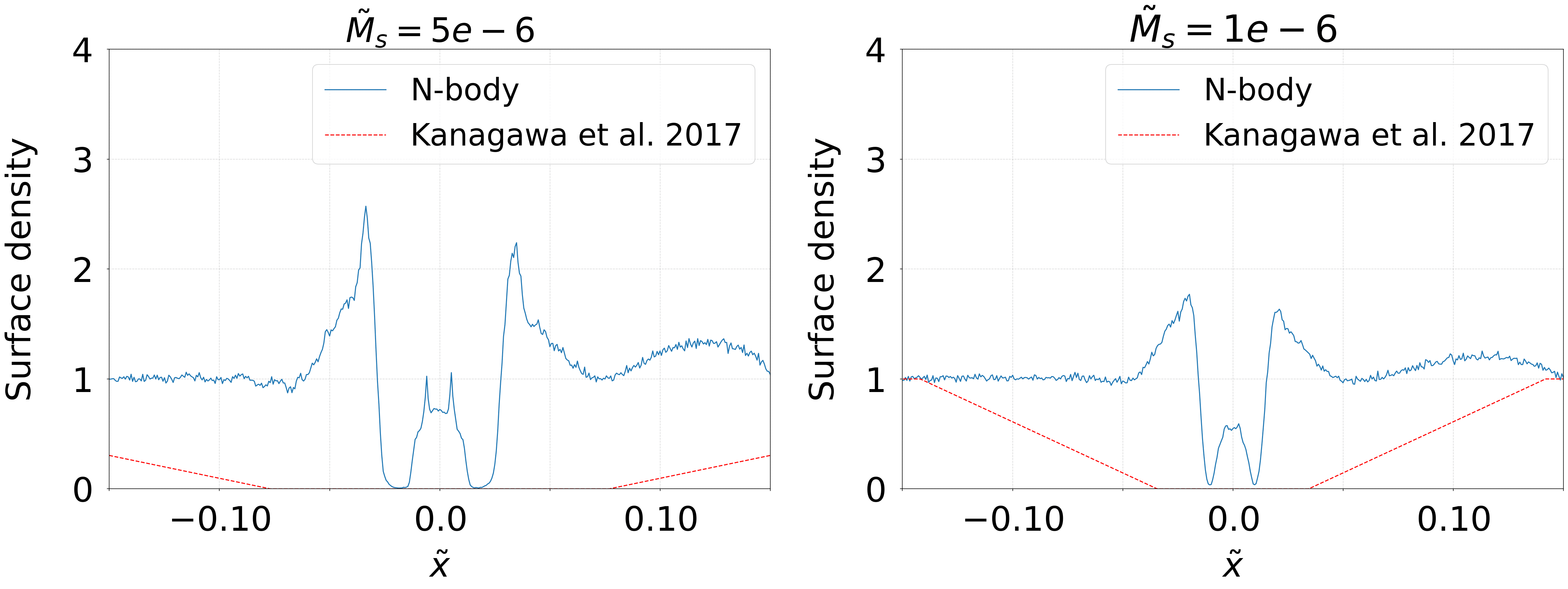}
\caption{The azimuthally-averaged surface density obtained in our simulation. The surface density is normalized by the initial surface density ($\tilde{t}=0$). Red dotted line is the gap profile based on \cite{Kanagawa:2017}'s formula. The left and right panels correspond to Model 1 ($\tilde{M}_{\rm s}=5\times10^{-6}$) and Model 2 ($\tilde{M}_{\rm s}=1\times10^{-6}$), respectively.}
\label{fig:sur_den}
\end{figure*}

Figure~\ref{fig:sur_den} shows the radial gap profiles obtained by azimuthal averaging of the surface density in Model 1 and Model 2. 
The surface density is normalized by the initial one. 
For comparison, the gap profile predicted for a gas disk by
\cite{Kanagawa:2017}'s formula is also plotted as red dotted lines. 
While \cite{Kanagawa:2017} assumed a constant $\alpha$ viscosity, we substitute the viscosity relevant to planetary rings into the \cite{Kanagawa:2017}'s formula (see Appendix A).
Compared with \cite{Kanagawa:2017}'s formula, extremely sharp gap edges are created in the N-body simulation. 
The sharp edges are more pronounced because the surface density is enhanced at the gap edges. 
This is caused by the negative diffusion discussed in Section~\ref{DensityWave}.
On a longer timescale, the pile-up could diffuse to some extent, and the negative diffusion would balance with the normal diffusion to be a steady state still with the pile-up (see below).
Actually, the enhanced optical depth at the edges was also pointed out by \cite{Showalter:1986} in the Voyager occultation data and by local N-body simulation \citep[e.g.,][]{Lewis:2000, Lewis:2011}. 
The two spike structures exist in the vicinity of the satellite orbit in \Figref{fig:sur_den}. 
This feature is created by clustering particles around the satellite, which is discussed in \Secref{Clustering}.

Angular momentum transport in rings is evaluated using the $r\theta$ component of the pressure tensor defined as
\begin{equation}
p_{r\theta}=\sum_{i}(v_{ir}-u_{r})(v_{i\theta} - u_{\theta}),
\end{equation}
where $v_{ij}$ is the velocity of the $i$-th particle in the $j$-direction and $u_{j}$ is the mean local velocity in the $j$-direction ($j = r$ or $\theta$) \citep{Lewis:2000}. 
It is a momentum transport rate in a $r$-direction across the plane perpendicular to the $\theta$-direction. 
The angular momentum flux is evaluated by multiplying the $p_{r\theta}$ by the semi-major axis of the considered area. 

In unperturbed rings, the angular momentum flux is directed outward, thus the angular momentum luminosity, which is time-integrated angular momentum flux, is also outward. 
In this case, the ring spreads outward while most materials move inward. 
However, \cite{Borderies:1989} suggested that the angular momentum luminosity reverses when the streamlines of particles are strongly perturbed, and it may sharpen gap edges. 
This requires that the sign of the time-integrated value of $p_{r\theta}$ changes at the gap edge.
We test this argument with our simulation results. 
With the same grids and time average as \Figref{fig:heatmap}, we evaluate $p_{r\theta}$ on the $\tilde{x}$-$\theta$ plane, where we assume that the velocity field is stationary and calculate $u_{\alpha}$ by time-averaging the local velocity at each grid. 

\begin{figure}[h]
\centering
\includegraphics[width=0.6\linewidth]{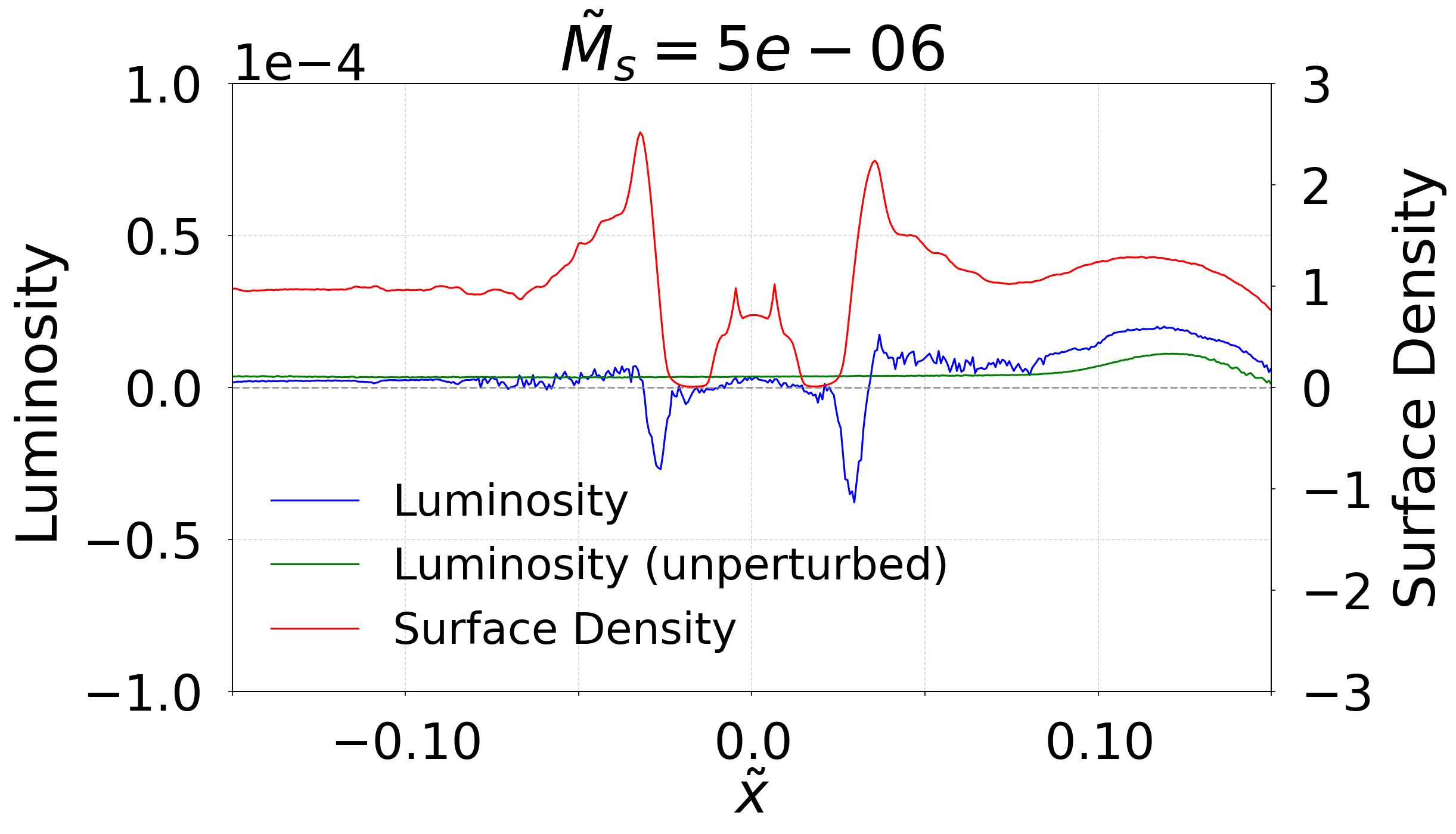}
\caption{The azimuthally-averaged angular momentum luminosity (the blue curve) and surface density (the red curve). For comparison, we also plot the unperturbed luminosity by the green curve.}
\label{fig:luminosity}
\end{figure}

\begin{figure}[h]
\centering
\includegraphics[width=0.8\linewidth]{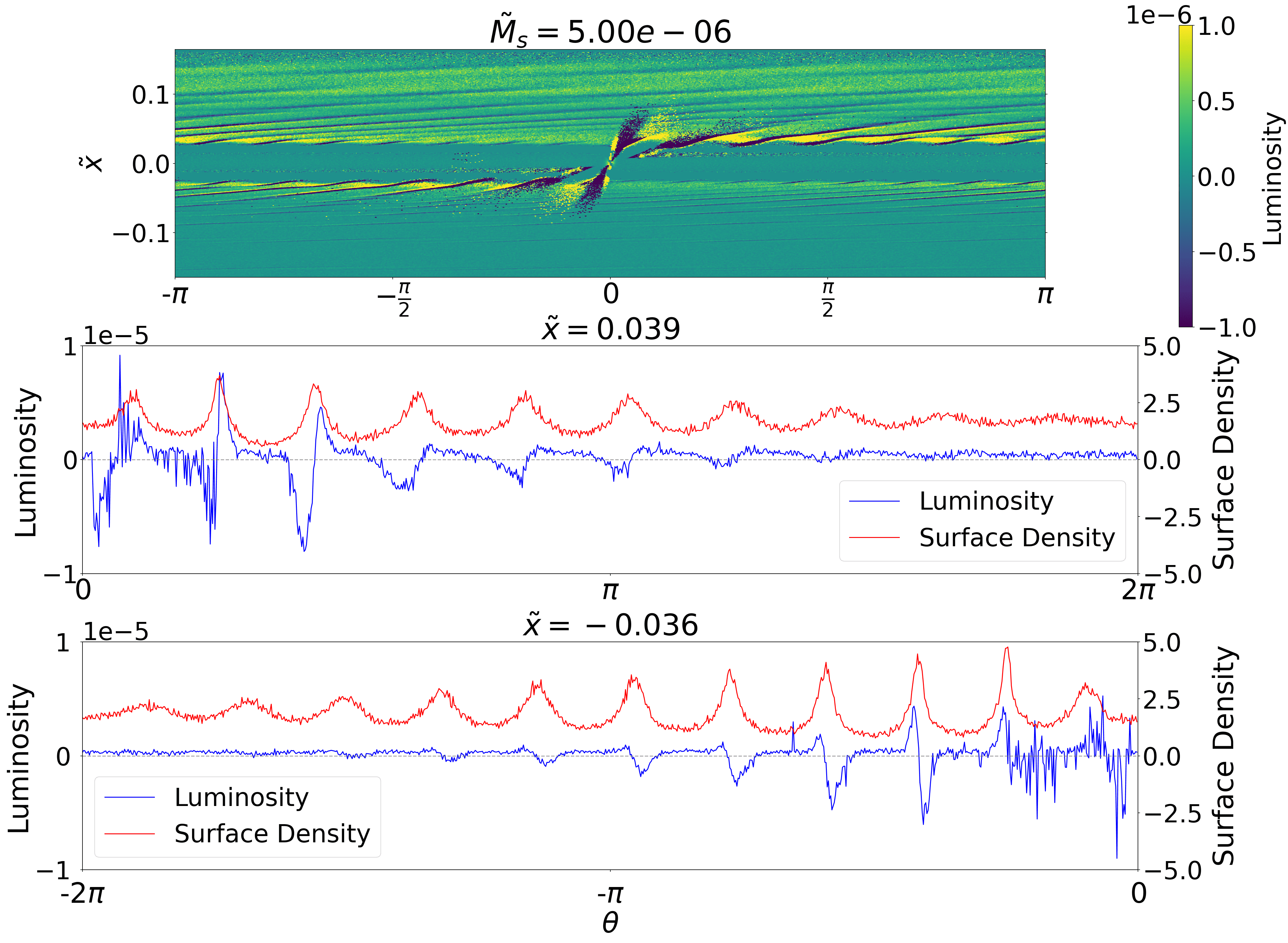}
\caption{The angular momentum luminosity  is plotted in the $\tilde{x}$-$\theta$ plane for Model 1 in the upper panel. 
The cross-sections of the surface density (the red curve) and the angular momentum luminosity (the blue curve) at $\tilde{x}= 0.039$ and $-0.036$ are shown as a function of $\theta$ in the middle and lower panels. 
}
\label{fig:heatmap_prt_5e-6}
\end{figure}

\begin{figure}[h]
\centering
\includegraphics[width=0.8\linewidth]{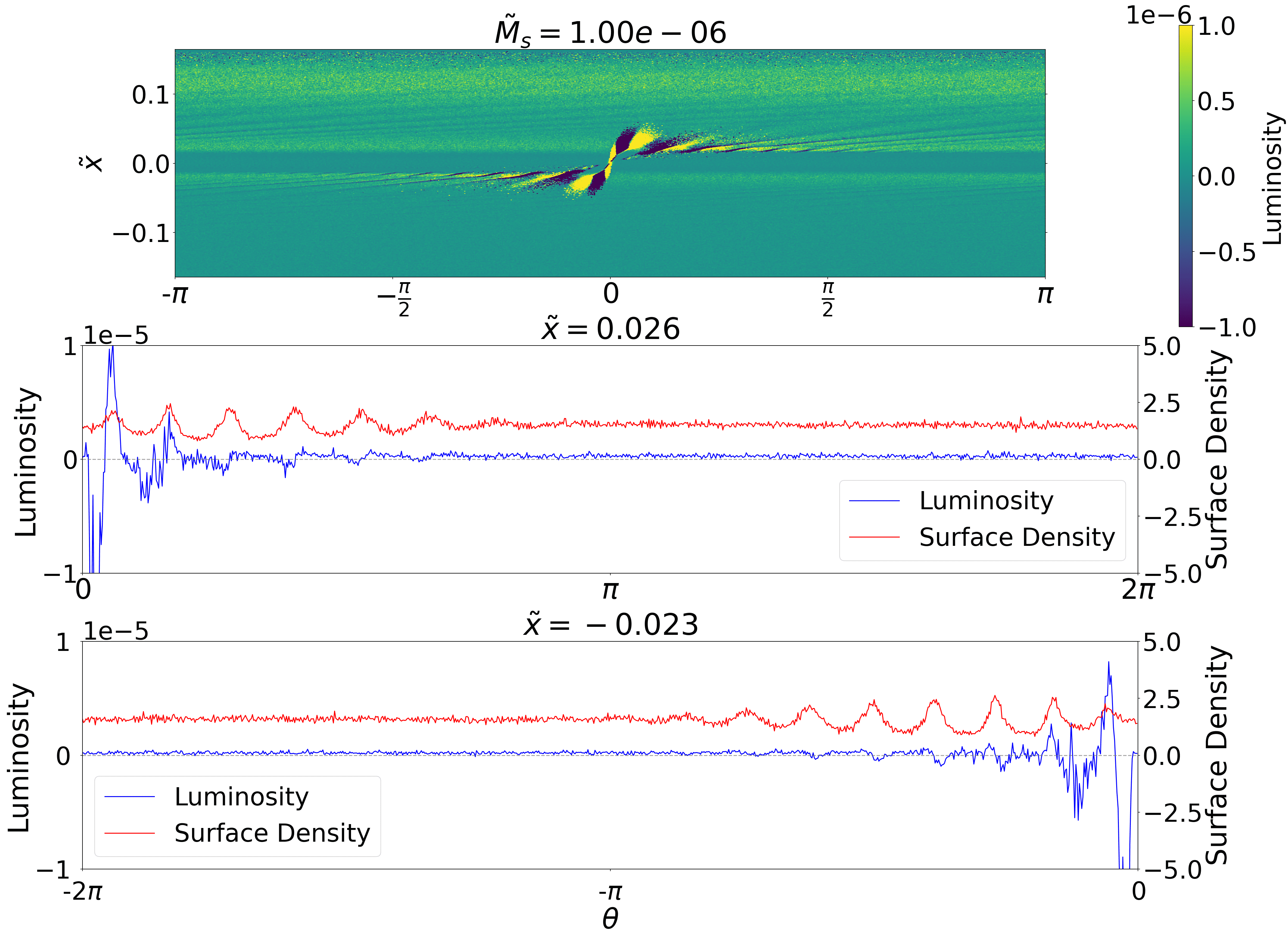}
\caption{Same as \Figref{fig:heatmap_prt_5e-6} but for Model 2.}
\label{fig:heatmap_prt_1e-6}
\end{figure}

Figure~\ref{fig:luminosity} shows the azimuthally averaged luminosity of annulus as a function of $\tilde{x}$ (blue curve) and the surface density normalized by the initial value (red curve). 
The luminosity reversal near the gap edge is clearly confirmed: it is positive outside the edge of the gap, but rapidly changes the sign and becomes negative at the edge of the gap. 
For comparison, we also conduct an additional run without the satellite and plot the unperturbed luminosity as a green curve. 

The $\tilde{x}$-$\theta$ distribution of the luminosity is shown in the upper panels of \Figref{fig:heatmap_prt_5e-6} and \Figref{fig:heatmap_prt_1e-6} with cross-sections along a horizontal curve in the vicinity of the edge.
On the density wavefronts, the particle trajectories are strongly perturbed and the surface density has strong peaks. 
The sign of the luminosity is reversed at the same point of the surface density peak, which is consistent with the argument of \cite{Borderies:1989}. 
In the downstream from the encounter, the reversal decays as the density waves decay and the surface density peaks become low.
In this region, the normal positive diffusion tends to smooth out the sharp edge. 
However, during a synodic period of the satellite and a particle in the gap edge ($\tilde{t}_{{\rm{syn}}}=2\pi/|\tilde{\Omega}_{\rm s} - \tilde{\Omega}| \sim 76$), the diffusion length is $\tilde{r}_{{\rm{diff}}}=\sqrt{\tilde{\nu} \tilde{t}_{{\rm{syn}}}}\sim6\times10^{-4}$, where we evaluate the viscosity $\nu$ with \Eqref{eq:viscosity_thin} and $\Omega_{\rm s}$ is the angular velocity of the satellite. 
Because $\tilde{r}_{{\rm{diff}}}$ is about two orders of magnitude smaller than the gap width $\sim0.04$, particles experience the next encounter before the sharp gap edges are largely smoothed, resulting in preserving the sharp gap edges.
This is also the case for the Keeler and Encke gaps. 
The diffusion length during the next encounter is $\sim0.7 \ {\rm{km}}$ for the Keeler gap and $\sim0.4 \ {\rm{km}}$ for the Encke gap with $\nu\sim20{\rm{cm^2/s}}$ for the Keeler gap and $\nu\sim70{\rm{cm^2/s}}$ for the Encke gap, respectively, where we use the viscosity values estimated by \cite{Gratz:2019} based on their 1D diffusion model. 
These lengths are sufficiently small compared to the gap width (see \Tabref{tab:value_summary}).

\subsection{Gap width}\label{GapWidth}

\begin{figure*}[h]
\centering
\includegraphics[width=\linewidth]{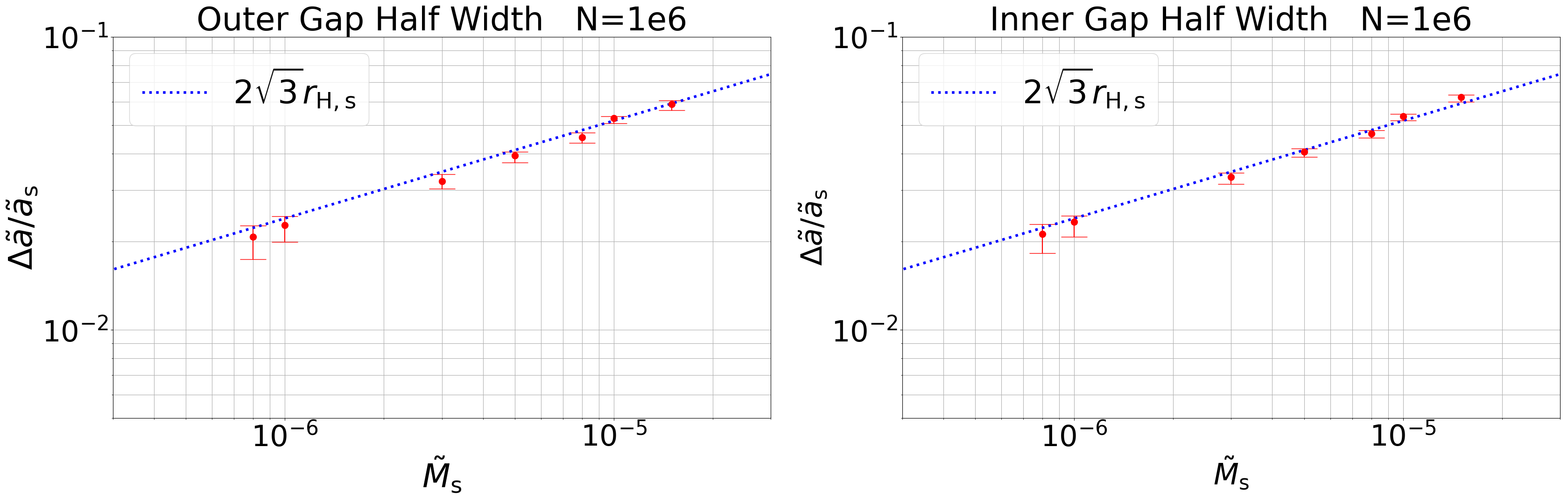}
\caption{The gap width is shown as a function of the satellite mass (Model 3-9). Blue dotted line represents a line proportional to one-third power of the satellite mass: $\Delta a \sim 2\sqrt{3} \, r_{\rm H,s}$. Left and right panel correspond to the outer and inner edges. We used $0.8\Sigma_0$ as the threshold of the surface density to define the gap width. The gap width in the case of upper threshold value ($1.0\Sigma_0$) and lower threshold value ($0.5\Sigma_0$) is also shown as error bars of each plot.}
\label{fig:gap_width}
\end{figure*}

Because the gap edges in the rings are extremely sharp with pile-ups, the gap half width is clearly defined.
Figure~\ref{fig:gap_width} shows the gap half width as a function of the satellite mass in our N-body simulation results with $N=1\times10^6$ (Model 3-9). 
In this figure, the threshold value to define the gap edge is $\Sigma_{\rm gap} = 0.8\Sigma_0$ with the error bars of the width with the threshold value from $\Sigma_{\rm gap} = 0.5\Sigma_0$ to $\Sigma_{\rm gap} =1.0\Sigma_0$.
The results show that the gap half widths $\Delta a$ is scaled by the satellite Hill radius ($r_{\rm H,s}\propto M_{\rm s}^{1/3}$): 
\begin{equation}
    \Delta a \sim 2\sqrt{3} \, r_{\rm H,s}. 
    \label{eq:gap_width}
\end{equation}
This gap width corresponds to the minimum impact parameter for the Jacobi integral of a particle in a circular orbit to be negative. 
The Jacobi integral $E$ is given by
\begin{equation}
    E=\frac{1}{2}(e_{\rm H}^2+i_{\rm H}^2)-\frac{3}{8}b_{\rm H}^2-\frac{3}{(r/r_{\rm H,s})}+\frac{9}{2},
\end{equation}
where $r$ is the distance between the particle and the satellite \citep{Ida:1989}. 
Particles in a circular orbit with $E<0$ (i.e. $b>2\sqrt{3}r_{\rm H, s}$) cannot enter the satellite's Hill sphere and never collide with it; in contrast, those with $E>0$ (i.e. $b<2\sqrt{3}r_{\rm H, s}$) collide with the satellite and create a cluster around it (see \Figref{fig:clustering}) or undergo a horseshoe turn, thus the gap width becomes $\sim2\sqrt{3}r_{\rm H, s}$. 
The error bars are wider and the gap edges are less sharp for smaller masses of the satellite. 

In general, the angular momentum transport due to the Lindblad resonances tends to open a gap. 
However, its contribution is not dominant in our simulation for the following reason, thus the gap width can be scaled only with the Hill radius of the satellite.
\cite{Goldreich:1980} calculated the torque density exerted by the embedded satellite on the ring considering the contribution from the high $m$-th Lindblad resonances, which is equivalent to weak gravitational scattering of distant encounters without orbit crossing. 
The total torque on the ring is given by
\begin{equation}
    T^{\rm L}=-0.84r^4\Sigma\Omega^2\qty(\frac{r}{a_{\rm s}-r})\qty(\frac{M_{\rm s}}{M_{\rm p}})^2,
    \label{eq:torque}
\end{equation}
where $\Sigma$ is the surface density of the ring and $\Omega=\sqrt{GM_{\rm p}/r^3}$ is the Kepler angular velocity \citep{Goldreich:1982}.
The timescale of the evolution driven by the Lindblad torque can be written as
\begin{equation}
    t_{\rm L}^{-1}=\qty|\frac{1}{r}\dv{r}{t}|=1.68\frac{a_{\rm s}^2\Sigma\Omega M_{\rm s}}{M_{\rm p}^2}\qty|\frac{a_{\rm s}}{a_{\rm s}-r}|.
\end{equation}
In our simulation, the timescale $\tilde{t}_{\rm L}\sim2\times10^5$ is much longer than the duration of the simulation $\tilde{t}\sim1000$, thus its contribution is not effective and strong scatterings with close encounters almost determine the gap width by the satellite. 
An equilibrium point where the torques balance is obtained by equating the Lindblad torque density and the viscous torque density and the gap width is predicted to be proportional to $M_{\rm s}^{2/3}$ \citep[e.g.,][]{Lissauer:1981, Goldreich:1982}. 
This dependence is different from our result of $\propto r_{\rm H,s} \propto M_{\rm s}^{1/3}$ (\Figref{fig:gap_width}).
In the real system, the half-widths of the Keeler and Encke gap are $\sim 3.2r_{\rm H, Daphnis}$ and $\sim 8.8r_{\rm H, Pan}$, where $r_{\rm H,Daphnis}$ and $r_{\rm H,Pan}$ are the Hill radius of Daphnis and Pan, respectively (see Table 2).
The Keeler gap width is consistent with our estimate, although the Encke gap width is larger.
The latter case may be contributed to by the long time effect of Lindblad torque, because Pan is 50 times more massive than Daphnis.

\subsection{Vertical structure of the satellite wake}\label{Vertical}

\begin{figure*}[h]
\centering
\includegraphics[keepaspectratio, width=0.7\linewidth]{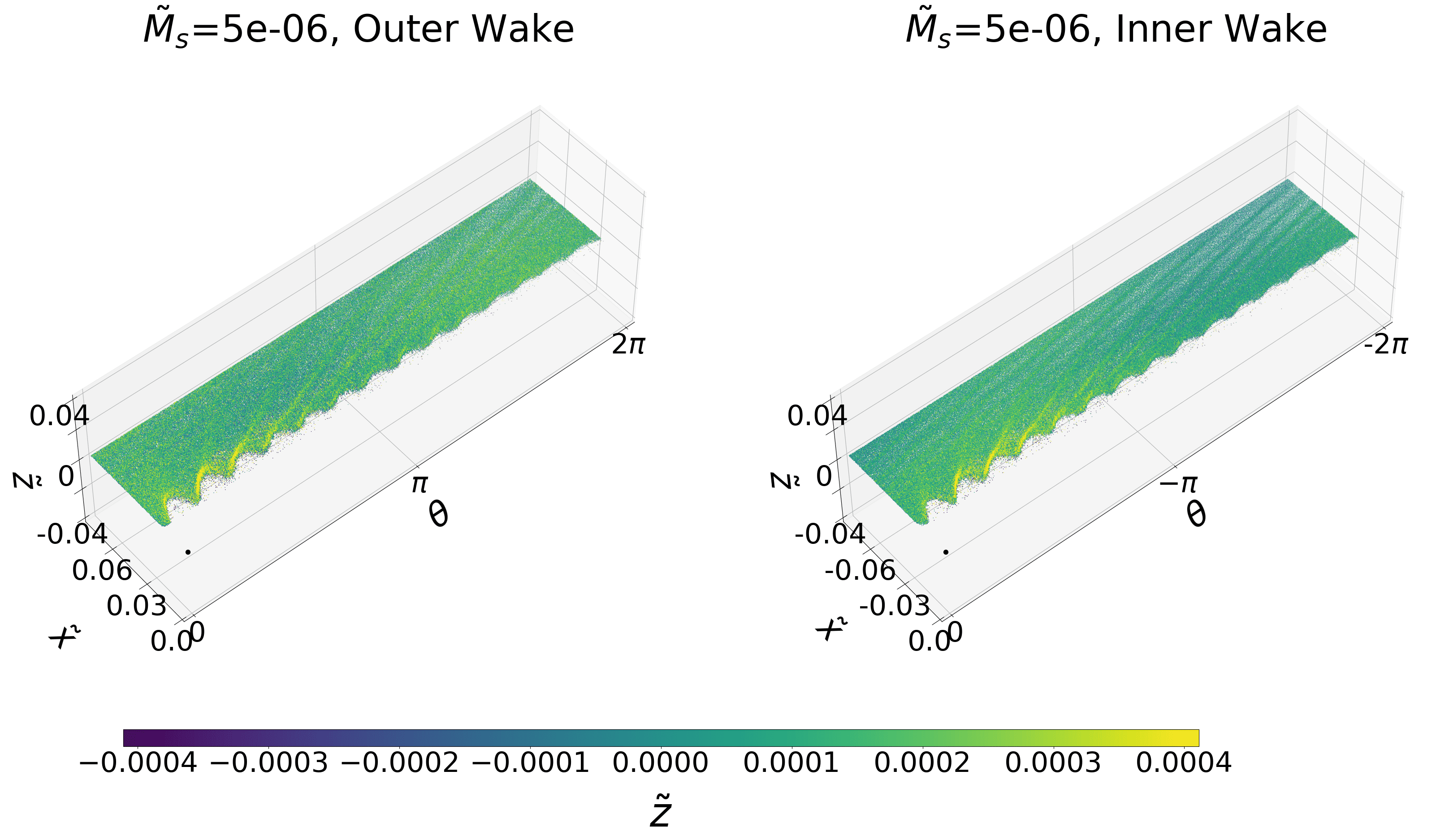}
\caption{The 3D plot of the inner and outer satellite wakes of Model 1. 
Color shows the $z$-component of each particle. 
Particles of yellow and dark blue create mountain-like vertical wall structures.}
\label{fig:3Dwake}
\end{figure*}

Figure~\ref{fig:3Dwake} shows the mountain-like vertical wall structure of the satellite wakes in inner and outer regions from the satellite orbit.
The $ z$-component of its position colors each particle. 
Yellow or dark blue particles correspond to vertically splashing particles. 

\begin{figure*}[h]
\centering
\includegraphics[keepaspectratio, width=0.9\linewidth]{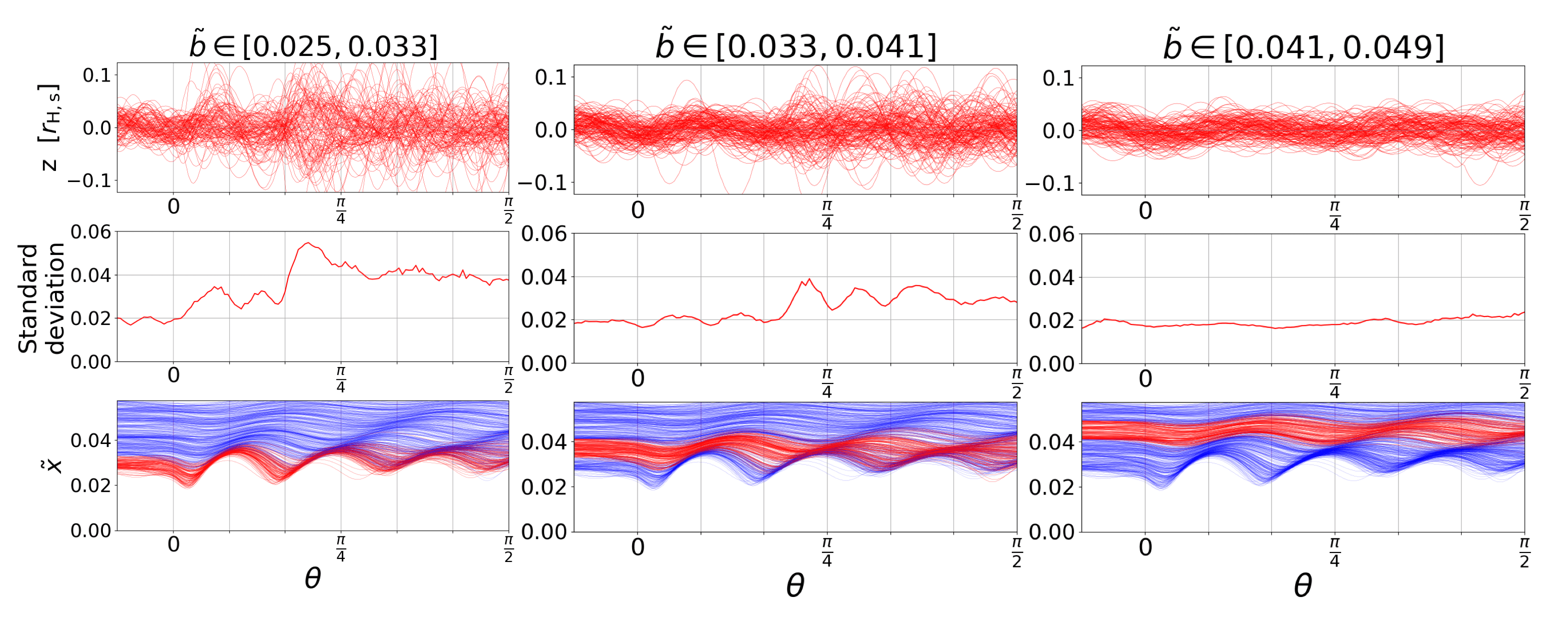}
\caption{The vertical excursion of particles (the upper panels)
and the standard deviation of the vertical spread (the middle panels) 
are plotted for the red-colored trajectories in the lower panels.
The vertical excursion of particles is shown in the unit of the satellite's Hill radius.}
\label{fig:streamline_height}
\end{figure*}

The vertical distributions of particles in the outer region in the unit of $r_{\rm H,s}$ and the standard deviation of the vertical spreads are shown in the upper and middle panels in \Figref{fig:streamline_height}.
The corresponding trajectories of particles are shown in the bottom panels (by the red curves).
The particle splashing is more pronounced as the impact parameter decreases. 
Near the wavefronts where the trajectories are strongly concentrated (e.g., $\pi/6<\theta<\pi/4$ in the left column), particles are strongly splashing and the standard deviations are significantly large. 
The splashing is the largest at the second wake, and its vertical height is as large as $\sim 0.1\, r_{\rm H,s}$. 
In the first wake, the epicyclic phase is strongly synchronized, thus the collisional velocity is relatively low (see also \Figref{fig:streamline}). 
In the second wake, the epicycle oscillation phase of particles start deviating from the synchronization due to the Kepler shear.
Accordingly, they undergo collisions with higher relative velocity and strongly splash in the vertical direction.

\begin{figure}[h]
\centering
\includegraphics[keepaspectratio, width=0.8\linewidth]{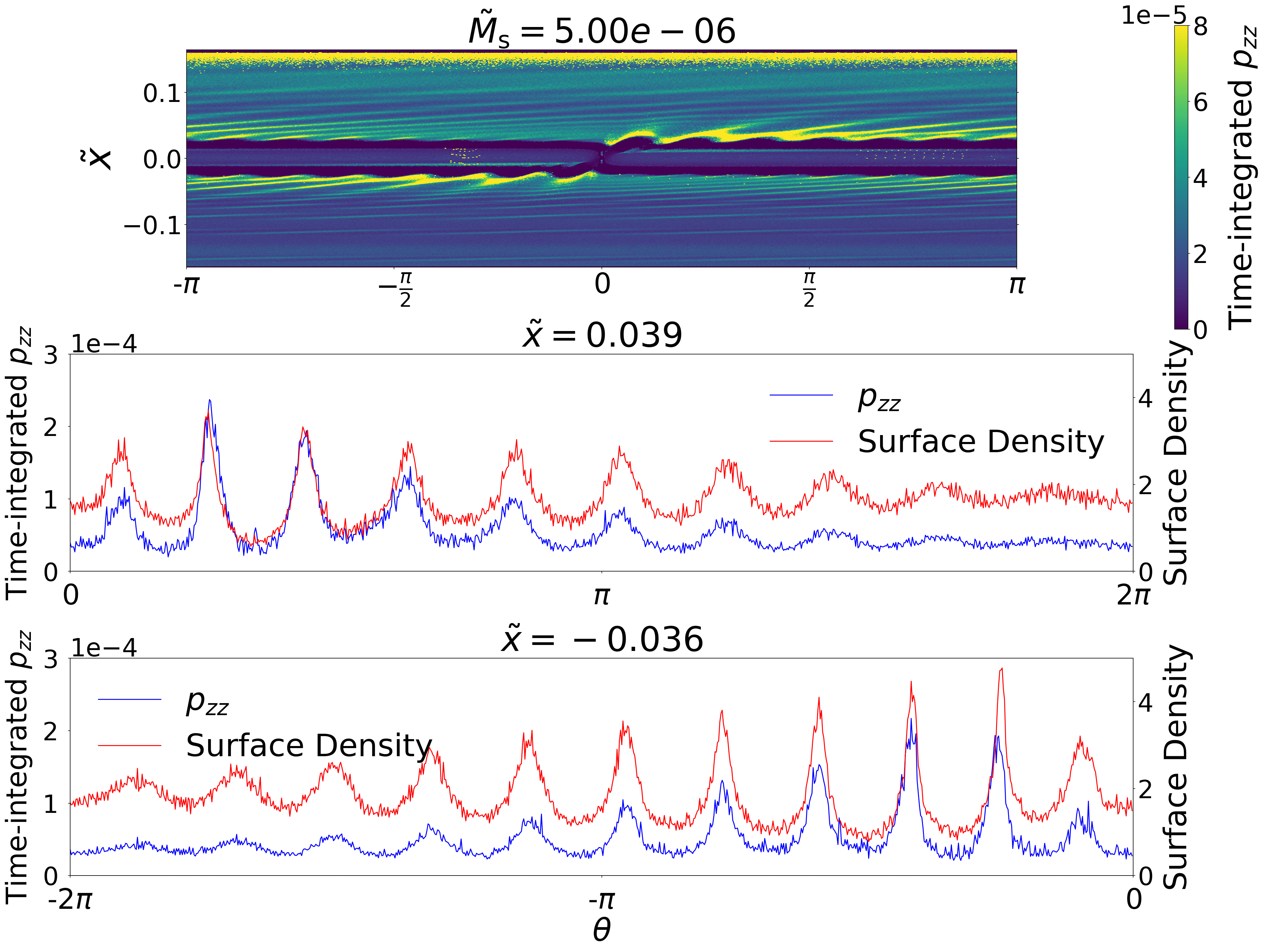}
\caption{The time-integrated $zz$-component of the pressure tensor in Model 1
in the $\tilde{x}-\theta$ plane (the upper panels),
and the cross-section at a specific $\tilde{x}$ (the middle and lower panels). The red and blue curves are the surface density and the time-integrated $zz$-component of the pressure tensor.}
\label{fig:heatmap_pzz_5e-6}
\end{figure}

\begin{figure}[h]
\centering
\includegraphics[keepaspectratio, width=0.8\linewidth]{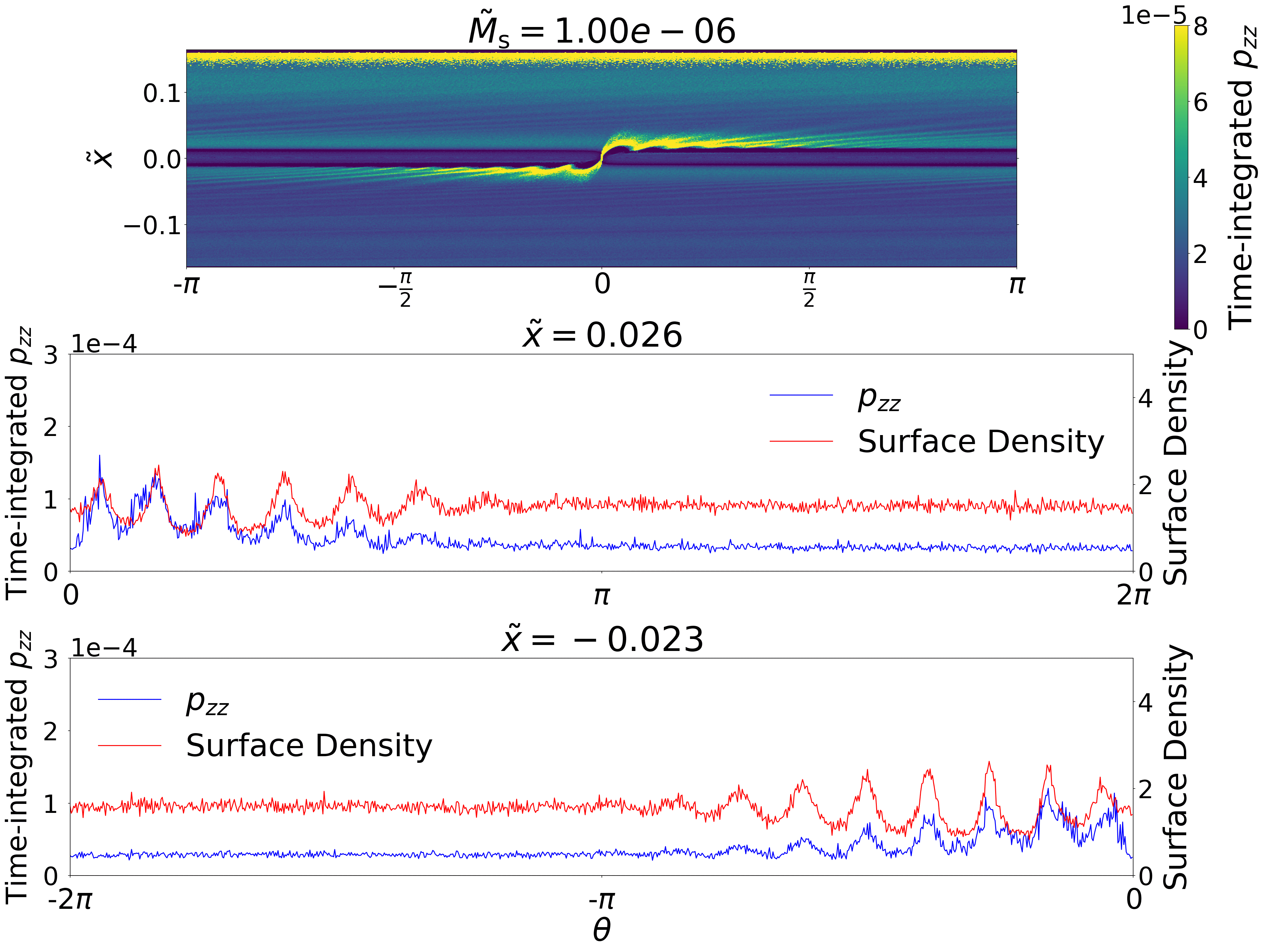}
\caption{Same as \Figref{fig:heatmap_pzz_5e-6} but for Model 2.}
\label{fig:heatmap_pzz_1e-6}
\end{figure}

The time-integrated $zz$-component of the pressure tensor is plotted in the $\tilde{x}$-$\theta$ plane in the upper panels in \Figref{fig:heatmap_pzz_5e-6} and \Figref{fig:heatmap_pzz_1e-6}.
The cross-sections at $\tilde{x} = 0.026$ and $-0.023$ are shown in the middle and lower panels. 
The integrated $p_{zz}$ and the surface density have peaks at the same locations. 
They decay together in the downstream.
These plots also show that the particle splashing is the largest in the second peaks. 

\begin{figure}[h]
\centering
\includegraphics[keepaspectratio, width=0.8\linewidth]{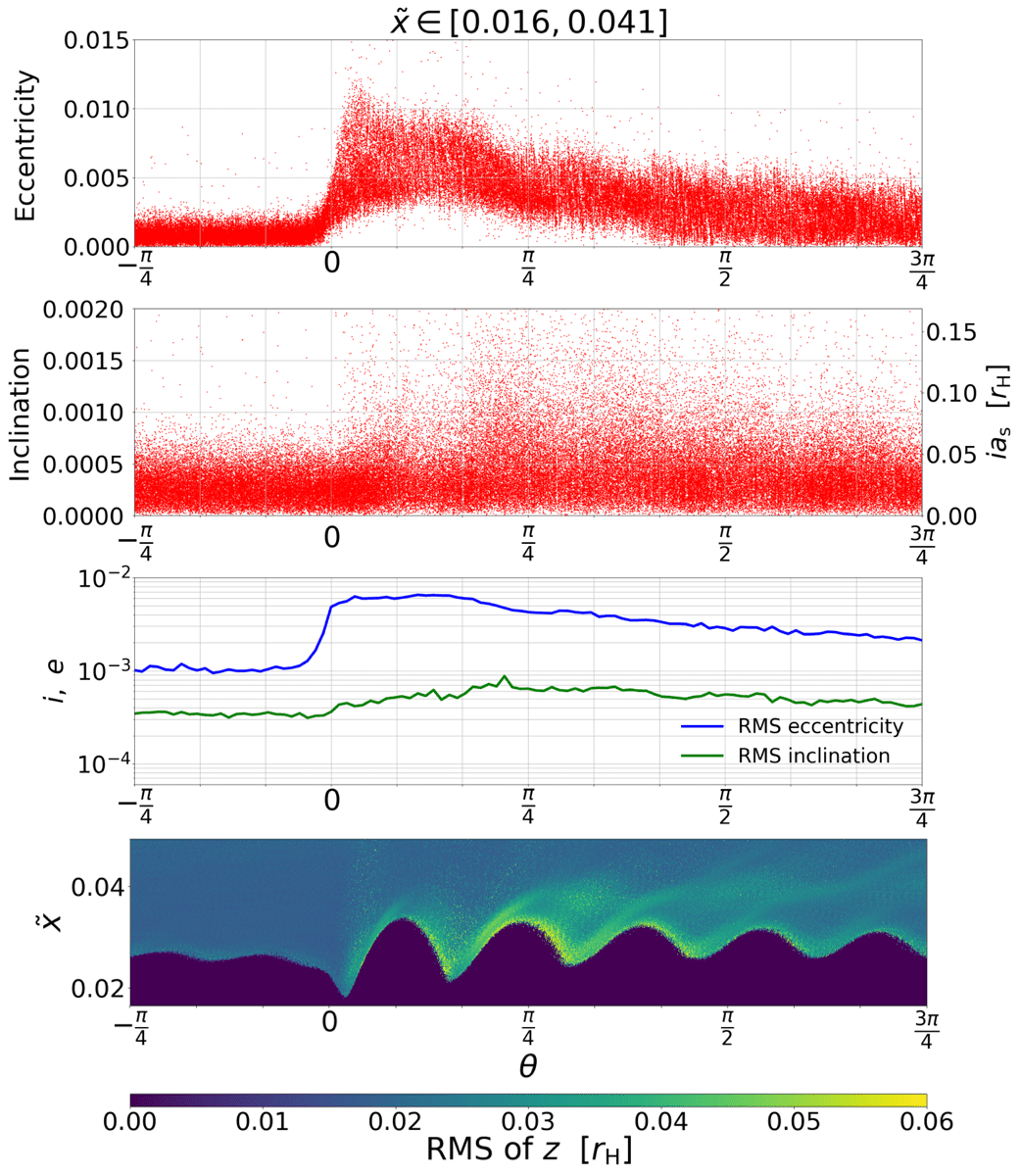}
\caption{The eccentricity (top panel) and inclination (second panel) of particles with the radial distance from the satellite orbit $\tilde{x}\in[0.016, 0.041]$ at $\tilde{t}=1000$ in Model 1. The right axis of the second panel shows the corresponding vertical height of particle splashing. The third panel from the top shows the root mean square (RMS) values of them. The bottom panel is a 2D color map of the root mean square of $z$ component of particle's position in the unit of Hill radius of the satellite.}
\label{fig:theta-ei}
\end{figure}

Here, we propose a new mechanism which naturally leads to the formation of the vertical mountain-like wall structures at the gap edge. 
The upper two panels of Figure~\ref{fig:theta-ei} shows the eccentricity and inclination of particles at the satellite wake region. 
We find that the excited eccentricity mostly damps during the second epicyclic phase after the encounter due to higher collisional velocity caused by deviation of the phase synchronization.
The inclination excitation phase is correlated to the eccentricity damping phase. 
The vertical height of the splashing is $\sim 0.1\,r_{\rm H,s}$ (the right axis of the second panel). 
This behavior is more clearly shown by the root mean square (RMS) values shown in the third panel.
These plots indicates that collisions convert the epicyclic motion of particles excited by the satellite's perturbation into its vertical motion, which is natural consequence of a collision between spherical particles. 
The bottom panel of Figure~\ref{fig:theta-ei} shows the $\tilde{x}$-$\theta$ map of root mean square values of $z$ component of the particle's position at each grid. 
The vertical structures are created along the wavefront of satellite wakes. 
They are more distinctive and taller at the second satellite wake than at the first (see also \Figref{fig:streamline_height}).
The created mountains is curved due to the Kepler shear.

Figure~\ref{fig:streamline_height} demonstrates that a single or a few high-velocity collision(s) quicky pumps up vertical motion.
In the first wake, because the epicycle phases are synchronized, collisional velocity is low and the splashing is not pronounced, although the collision probability is high in the first wavefront where the particle density is the highest.  
In the second wake, while the epicycle motions have not been damped yet and the particle density is kept high at the wavefront, the particle motions deviate from the synchronization.
This results in higher collisional velocity and pronounced vertical splashing at the wavefront.

The enhanced vertical thickness around the propeller structures may also be explained by a related process.
\citet{Hoffmann:2015} proposed that the lateral ``thermal'' motions excited by the fully embedded satellite in the propeller structure would be converted to vertical thermal motions through diffusion due to many times weak collisions to establish a quasi-equilibrium.
In our case of a clear-gap-opening satellite, the conversion occurs through a single or a few high-velocity collision(s) and creates the striking vertical wall structures at the second wavefront.  

Applying our model, we estimate the height of observed vertical structures at Keeler and Encke gaps.
The half-width of the Keeler and Encke gaps normalized by the Hill radius of the embedded satellites (Daphnis and Pan) is $\sim3.0$ and $\sim9.0$ (see \Tabref{tab:value_summary}), and the corresponding excited eccentricity normalized by $r_{\rm{H, s}}/a_{\rm s}$ is $\sim1.1$ and $\sim0.09$ based on \Eqref{eq:weiss}, respectively. 
Using typical results of our simulations, we adopt the collisional velocity $\sim 0.5$ times the excited epicycle velocity and $\sim 1/3$ of the conversion rate into the vertical component, the height of the splashing particles is estimated to be $\sim 0.18 \, r_{\rm H, Daphnis} \sim 850 \, \rm{m}$ at Keeler gap and $\sim 0.015\, r_{\rm H, Pan}\sim 274 \, \rm{m}$ at Encke gap, where $r_{\rm H, Daphnis}$ and $r_{\rm H, Pan}$ are the Hill radius of Daphnis and Pan, respectively. 
The lower value at Encke gap originates from the wider gap width around Pan.
The estimated height at the Keeler gap edge is consistent with the observational estimate, $\sim 1 \, \rm{km}$, from the shadow cast on the ring \citep{Weiss:2009}.
This theoretical estimate predicts that the vertical wall height at the Encke gap edge is $1/3$ of the Keeler gap.

\cite{Weiss:2009} argued that the vertical splashing on the Keeler gap edge is caused by the out-of-plane perturbations from Daphnis in the inclined orbit. 
Our trial N-body simulation with the inclined orbit of the embedded satellite shows similar results in vertical splashing as well as satellite wake patterns, sharp edges, and gap width, to those in the case with the circular and coplanar orbit. 
More detailed investigation on the effect of the inclination of the satellite is left for the future works (see also \Secref{Conclusion}).

\subsection{Effect of strongly excited  self-gravity wakes}\label{sec:StrongGravityWake}

While we have so far shown the results with $\tau_0 \sim 0.1$, the actual A ring around Saturn has $\tau_0 \sim 0.5$ \citep{Colwell:2009}. 
Although in general, high-resolution global N-body simulation becomes increasingly harder as the initial optical depth $\tau_0$ increases, we also carry out the simulations with $\tau_0 \sim 0.3$ (Model 10) and $\tau_0 \sim 0.5$ (Model 11). 
Using $\tau_0 = \pi (\Sigma/m) R^2$ and 
$\Sigma=M_{\rm ring}/S$, where $S$ is the surface area of the ring,
Toomore's $Q$ value (\Eqref{eq:Q} and \Eqref{eq:sigma}) is reduced to:
\begin{align}
    Q&=\frac{\Omega}{3.36G\Sigma}\sqrt{\frac{Gm}{R}}\notag\\
        &\simeq \frac{\pi^{1/4}}{3.36} \tau_0^{-1/4} 
        \tilde{m}^{1/4} \ 
        \tilde{r} \ ^{-3/2} \ 
        \qty(\frac{M_{\rm ring}}{S})^{-3/4}.
\end{align}
In Model 10 and Model 11, Toomore's $Q$ value is $\sim1.4$ and $\sim0.8$ at the satellite orbit ($r=0.66r_{\rm R}$), respectively, and self-gravity wakes develop much more than the models with $\tau_0 \sim 0.1$ ($Q\sim 4$).

\begin{figure*}[h]
\centering
\includegraphics[width=0.7\linewidth]{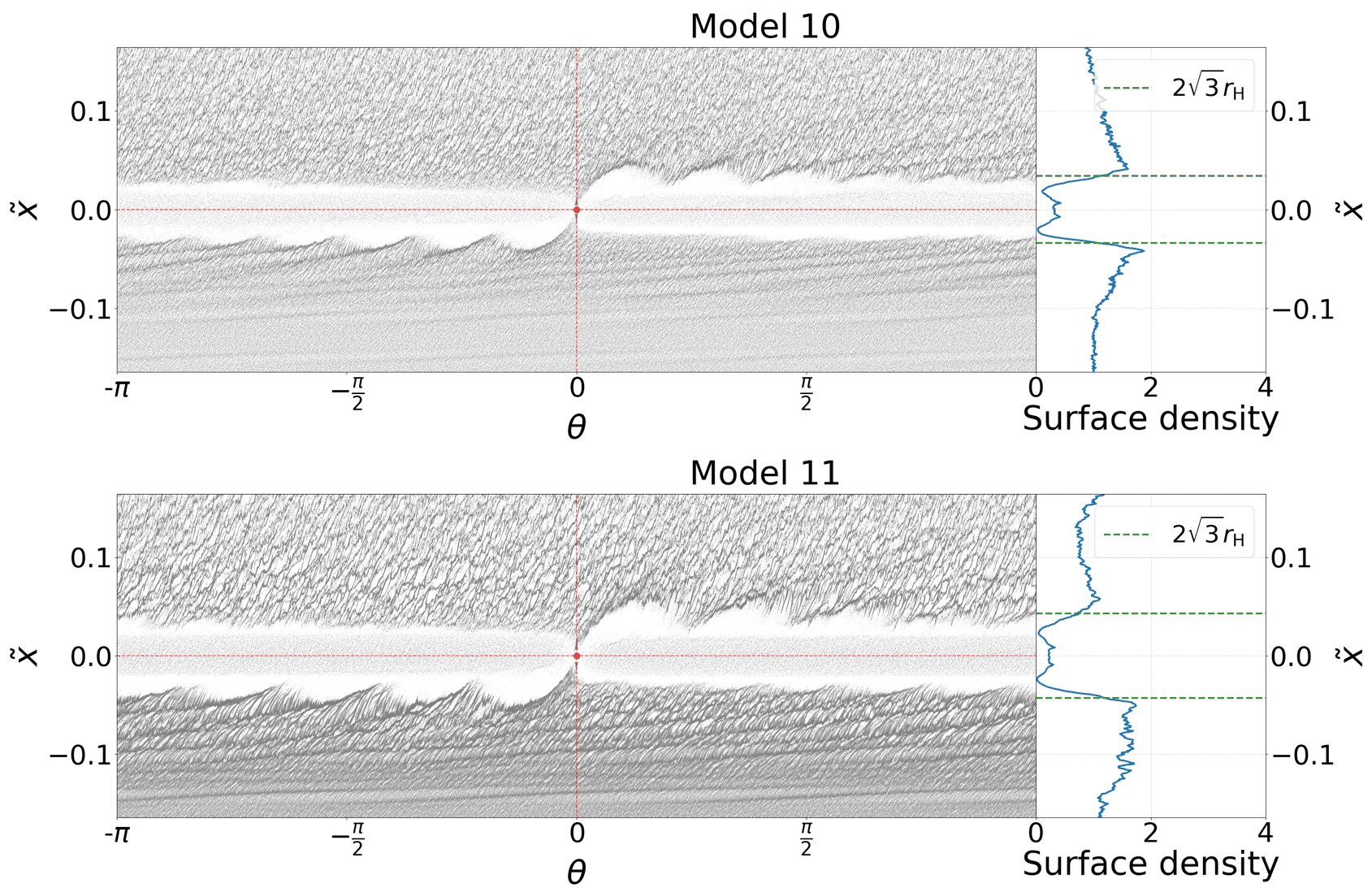}
\caption{The snapshot of created satellite wake and self-gravity wake (left) and surface density profile (right) at $\tilde{t}=1000$. Upper panel corresponds to Model 10 and lower panel corresponds to Model 11. The green dashed lines in the left panel shows the estimated gap width (see \Eqref{eq:gap_width}).}
\label{fig:wake_HighTau}
\end{figure*}

\begin{figure}[h]
\centering
\includegraphics[width=0.7\linewidth]{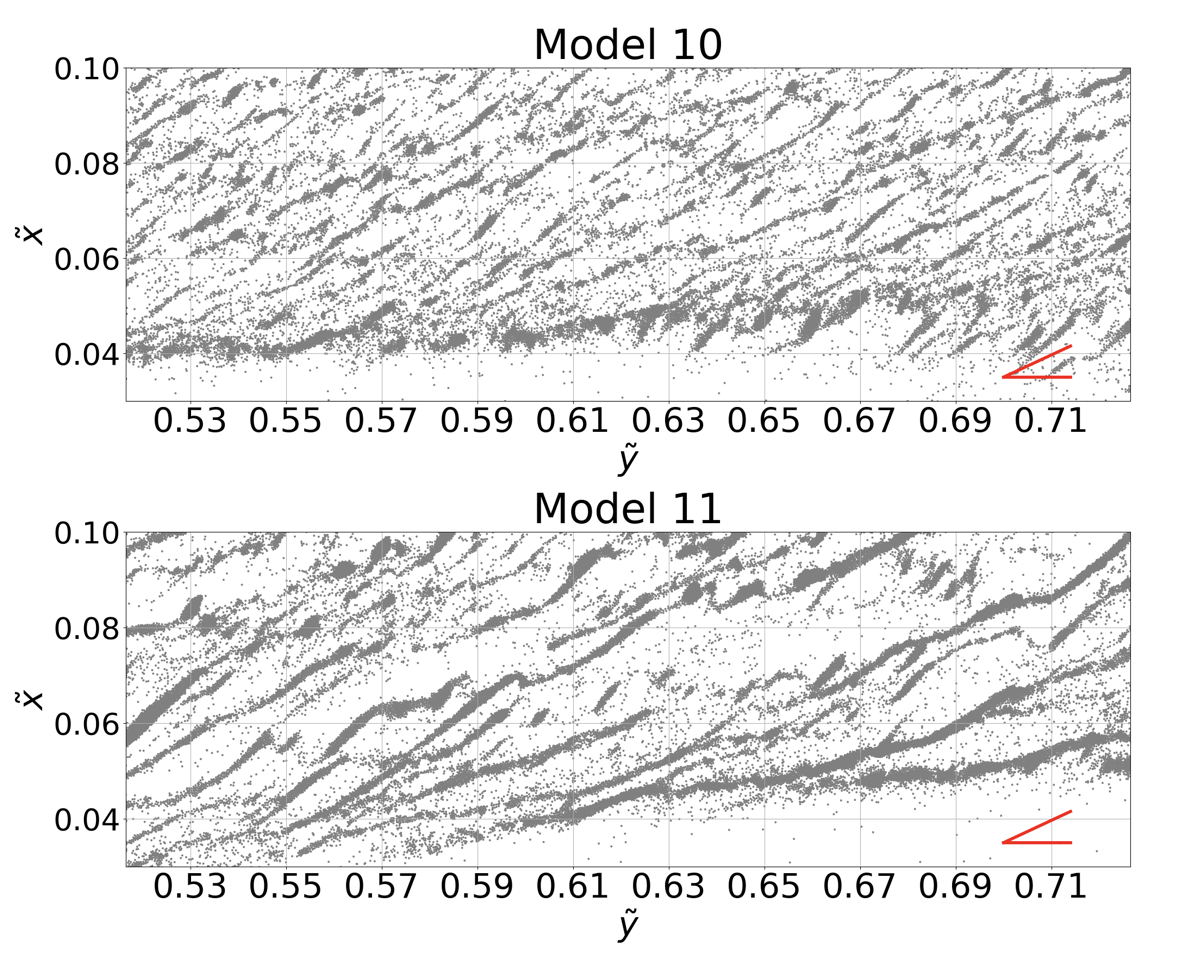}
\caption{The zoom-in view of the self-gravity wakes with a same scale of vertical and horizontal axis in Model 10 (upper panel) and Model 11 (lower panel). The red symbol of the lower left of each panel represents an estimated pitch angle ($\theta_{\rm p}\sim 25^{\circ}$) in local N-body simulation \citep[e.g.,][]{Salo:1995, Michikoshi:2015}.}
\label{fig:wake_highTau_zoom}
\end{figure}

\begin{figure*}[h]
\centering
\includegraphics[width=0.7\linewidth]{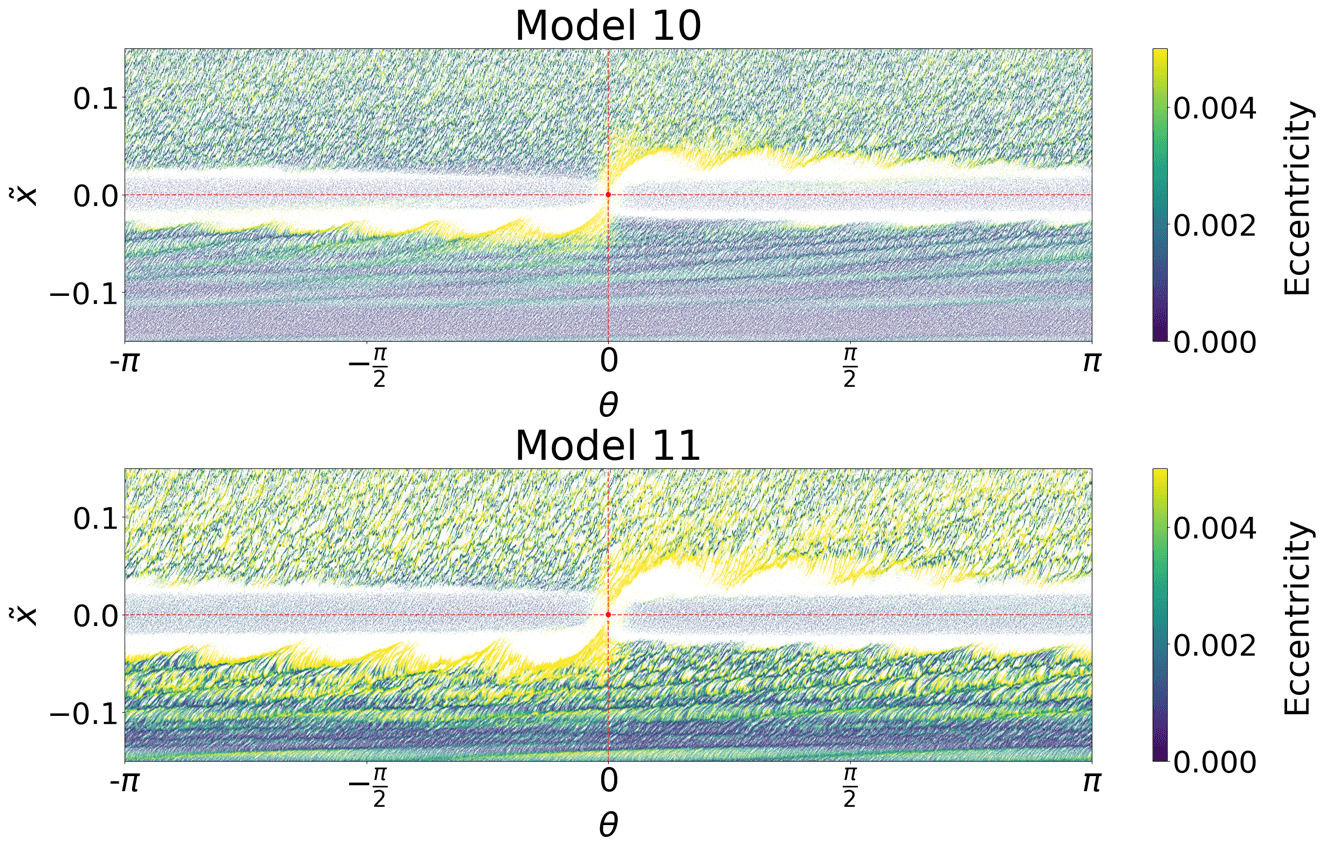}
\caption{The snapshot with particle's eccentricity shown as a color. Upper panel corresponds to Model 10 and lower panel corresponds to Model 11.}
\label{fig:wake_ecc_HighTau}
\end{figure*}

\begin{figure*}[h]
\centering
\includegraphics[width=0.7\linewidth]{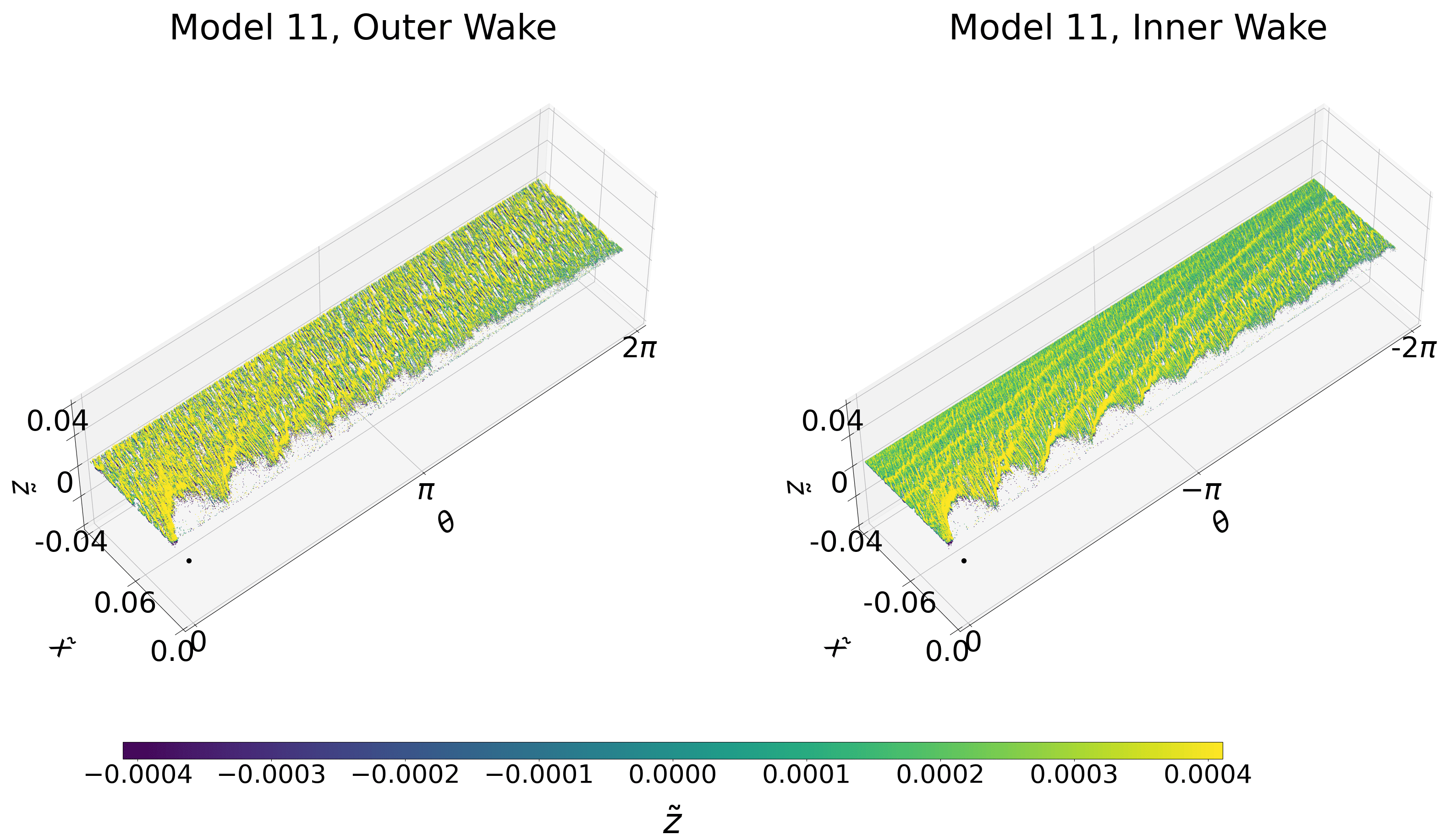}
\caption{3D plot of the inner and outer satellite wakes of Model 11. Color shows the $z$-component of each particle. }
\label{fig:3Dwake_HighTau}
\end{figure*}

Figure~\ref{fig:wake_HighTau} shows the azimuthally stretched snapshots of Model 10 and 11. 
When $Q\lesssim2$, the disk becomes gravitationally unstable, and self-gravity wakes develop \citep[e.g.,][]{Salo:1992, Salo:1995, Daisaka:1999}. 
Model 10 is marginal for the development of the self-gravity wakes (the upper left panel in \Figref{fig:wake_HighTau} and the upper panel in \Figref{fig:wake_highTau_zoom}).
Because $Q$ is higher for the inner region, the self-gravity wakes are weaker there, and the satellite wakes are still clear, while they are almost drowned out in the outer region. 
In Model 11, $Q$ is further lower, and the strong self-gravity wakes develop in both inner and outer regions (the lower left \Figref{fig:wake_HighTau} and the lower panel in \Figref{fig:wake_highTau_zoom}). 
The pitch angle of the self-gravity wakes ($\theta_{\rm p}$) is nearly consistent with the results of previous local N-body simulation; $\theta_{\rm p}\sim25^{\circ}$ \citep[e.g.][see \Figref{fig:wake_highTau_zoom}]{Salo:1995, Michikoshi:2015}.
The radial distribution of the surface density is shown in the right panel of \Figref{fig:wake_HighTau}. The green dashed lines show the estimated gap width from \Eqref{eq:gap_width}, which is consistent with the case where the strongly excited self-gravity wakes exist.
When the self-gravity wakes develop, the particle's eccentricity is strongly excited by them (\Figref{fig:wake_ecc_HighTau}). 
The vertical splashing at the wavefronts of the satellite wakes is also smoothed out, in particular in the outer region with lower $Q$ (\Figref{fig:3Dwake_HighTau}).

\cite{Lewis:2005} found through local N-body simulation that satellite wakes also disrupt self-gravity wakes and that ropey structures, which was observed by Cassini and called ``straw''  \citep{Porco:2005}, could form by the interaction between the satellite wakes and the self-gravity wakes. 
In our simulation, a similar structure appears: many elongated secondary structures extend from the wavefronts of satellite wakes.

Thus, the self-gravity wakes reduce the prominent features of sharp gap edges and vertical splashing at the wavefronts of satellite wakes. 
However, the reduction would be negligible in the actual ring-satellite system around Saturn for the following reasons. 
For the marginal state of $Q \sim 2$, the wavelength near the Roche limit is given by the Toomre critical wavelength \citep{Salo:1995, Daisaka:1999}:
\begin{equation}
\lambda_{\rm SG} \sim \frac{4\pi^2 G \Sigma}{\Omega^2}  \sim \frac{\pi \Sigma r^2}{M_{\rm p}} r \sim 4\pi r_{\rm R}\tilde{M}_{\rm ring}. 
\label{eq:lambda}
\end{equation}
On the other hand, the wavelength of satellite wakes is $\lambda_{\rm SW} \sim r_{\rm H,s}$. 
In Model 10 and 11, $\lambda_{\rm SG} \sim 4\pi r_{\rm R}\tilde{M}_{\rm ring} \sim 6 \times 10^{-3} r_{\rm R}$ and $1\times 10^{-2} r_{\rm R}$, while $\lambda_{\rm SW} \sim r_{\rm H,s} \sim 2 \times 10^{-2} r_{\rm R}$.
Because $\lambda_{\rm SG}$ is as large as $\sim (3-5)\times 10^{-1} \lambda_{\rm SW}$, the effect of the self-gravity wakes is not neglected.
On the other hand, in the real Saturnian system, $\lambda_{\rm SG} \sim 4\pi r_{\rm R}\tilde{M}_{\rm ring} \sim 10^{-7}r_{\rm R}$ and $\lambda_{\rm SW} \sim r_{\rm H,s} \sim (3\times 10^{-5} - 10^{-4}) \, r_{\rm R}$ for Pan and Daphnis (Table~\ref{tab:value_summary}).
Because $\lambda_{\rm SG} \lesssim 10^{-3} \lambda_{\rm SW}$, the effect of the self-gravity wakes would be much smaller than that in Model 10 and 11, suggesting that the sharp gap edges and vertical splashing at the wavefronts of satellite wakes would not be diminished. 
On the other hand, the self-gravity wakes must be considered for the effective viscosity (angular momentum transfer), because it is more important than that by particle collisions.

\section{Discussion}\label{Discussion}
\subsection{Comparison with the real system}\label{Comparison}

In the global N-body simulation of ring systems, we need to use much larger particles than in the real system. 
Here, we check how it affects the results. 
As shown in the previous sections, the formation of the satellite wakes, sharp edges, and vertical wall structures are driven by physical collisions between particles (e.g., ``negative diffusion'') and the gravitational perturbations by the satellite. 
The latter is scaled with the Hill radius of the satellite. 
The created structures depend only weakly on the strength of the former.
Therefore, at least from this point of view, our simulation reflects the mechanism working in Saturn's ring and our results can be directly applied to the real system.

On the other hand, the larger size of the particles overestimates collisional viscosity.
We show here that the overestimation would not affect our results.
The dynamical quantities used in this evaluation are listed in \Tabref{tab:value_summary}. 
\begin{table*}[h]
 \caption{Dynamical parameters of the embedded satellites Daphnis and Pan and the Encke and Keeler gaps \citep[data from][]{Weiss:2009}: $a_{\rm s}$, $M_{\rm s}$, and $r_{\rm{H, s}}$ are the semi-major axis, mass, and the Hill radius of the satellite and $\Delta a$ is the half-width of the gap.}
 \centering
  \begin{tabular}{cccccc}
   \hline
    Gap & Satellite & $a_{\rm s}$ [km] & $M_{\rm s}$ [g] & $r_{\rm{H, s}}$ [km] & $\Delta a$ [km] \\
   \hline \hline
   Encke & Pan & 133,584  & $4.4\times10^{18}$ & $18.3$ & 161 \\
   Keeler & Daphnis & 136,506 &  $6.8\times10^{16}$ & $4.66$ & 13-20 (inner)\\
          &         &         &                     &    & 14-16 (outer)\\
   \hline
   \label{tab:value_summary}
  \end{tabular}
\end{table*}
\begin{table*}[h]
\caption{Comparison of simulated systems with the real system. 
 The ratio of collisional ($\nu_{\rm coll}$) and gravitational ($\nu_{\rm grav}$) viscosities is listed for the real system and each model.
 Here, in the estimation of the real system, we assume that a particle radius $r_{\rm p}=10$ [m], the internal density of a particle $\rho=0.5$ [g/cm$^3$], and the optical depth $\tau=0.5$ \citep{Colwell:2009}.} 
 \centering
  \begin{tabular}{cccccc}
   \hline
       &  $\nu_{\rm coll}/\nu_{\rm grav}$ \\
   \hline \hline
   Pan    & 0.033 \\
   Daphnis & 0.026  \\
    \hline \hline  
    Model 1,2  & 1.428 \\
    Model 3-9  & 1.570 \\
    Model 10 & 0.506 \\
    Model 11 & 0.302 \\
   \hline
   \label{tab:Comparison}
  \end{tabular}
\end{table*}
We calculate the ratio of the collisional viscosity $\nu_{\rm coll}$ (\Eqref{eq:coll}) and the self-gravitational viscosity $\nu_{\rm grav}$ (\Eqref{eq:grav}). 
The transitional viscosity $\nu_{\rm trans}$ is given by the sum of the other two viscosity components; $\nu_{\rm trans}\sim\nu_{\rm coll} + \nu_{\rm grav}$. When the pronounced self-gravity wakes appear (Model 10 and 11), $\nu_{\rm grav}$ dominates $\nu_{\rm coll}$, thus $\nu_{\rm trans}\sim\nu_{\rm grav}$. 
In contrast, when the self-gravity wakes are not strong (Model 1-9), $\nu_{\rm coll}$ dominates $\nu_{\rm grav}$, thus $\nu_{\rm trans}\sim\nu_{\rm coll}$ \citep{Daisaka:2001}.

For the simulation models, we substitute the mass and physical radius of the super-particles and the surface density into the formulas.
For the real system, we use a particle radius $R = 10$ m, the internal density of a particle $\rho=0.5 \,{\rm g/cm^3}$ and the optical depth $\tau=0.5$ \citep{Colwell:2009}. 
These values could not be completely exact and vary depending on the position in the ring, thus this is an order estimate.

The evaluated values of the ratio are summarized in the \Tabref{tab:Comparison}. 
The ratios $\nu_{\rm coll}/\nu_{\rm grav} \gtrsim 1$ in Model 1-9 and $\nu_{\rm coll}/\nu_{\rm grav} \lesssim 1$ in Model 10 and Model 11. 
Thus, the effects of the gravitational and collisional viscosities are comparable, although the collisional viscosity is overestimated by using a much larger particle size in our simulations compared to the real system. 
However, as mentioned above, the structures we are focusing on here are scaled with the Hill radius because these are created by the gravitational perturbations by the satellite coupled with inelastic collisions between the ring particles.
Thus the overestimation for $\nu_{\rm coll}$ would not cause significant effects on the results.

\subsection{Particle clustering around a satellite}\label{Clustering}

\begin{figure}[h]
\centering
\includegraphics[width=0.8\linewidth]{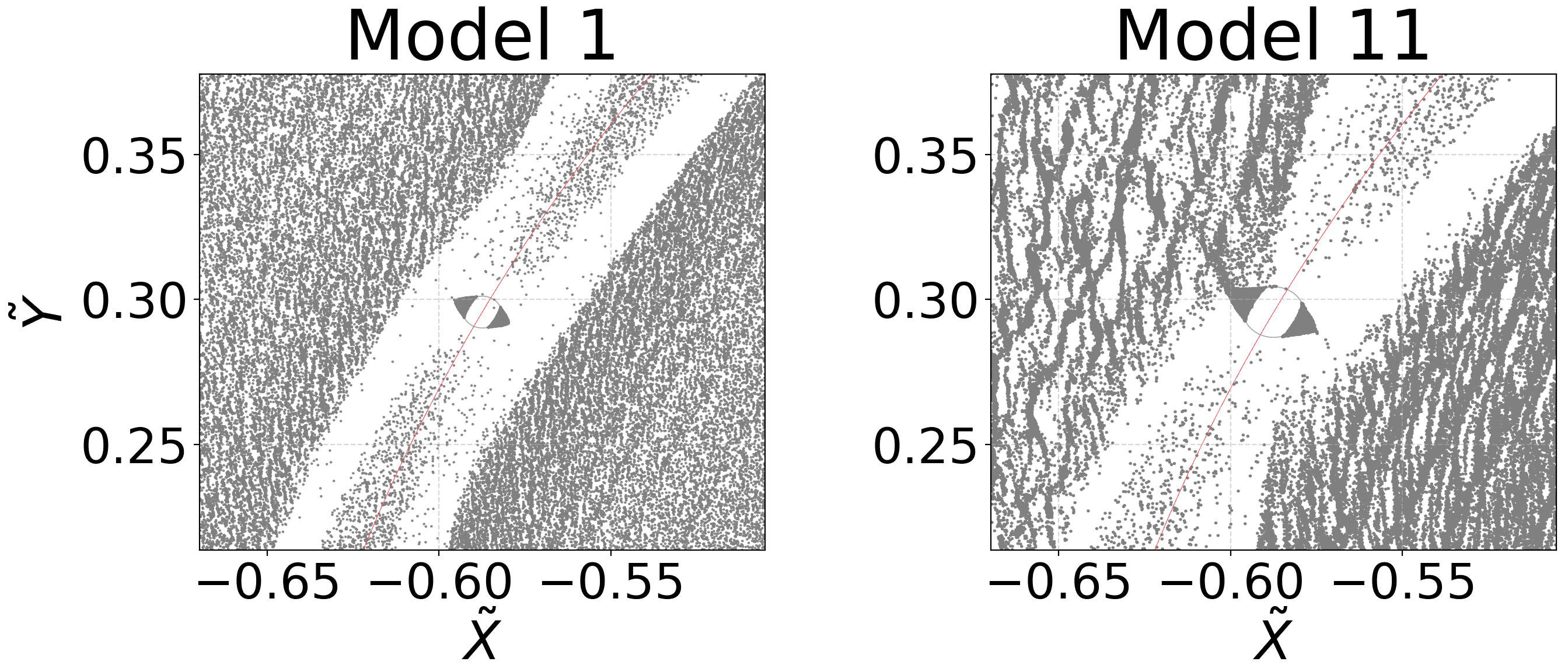}
\caption{Particles clustering around the satellite. 
The left and right panels correspond to Model 1 and Model 11, respectively. 
The particles and the satellite are shown with their actual size.
The red curve is the orbit of the satellite.}
\label{fig:clustering}
\end{figure}

Our simulation shows that the particles cluster around the satellite in a lemon shape as shown in zoom-in views in \Figref{fig:clustering}. 
Ring particles (the grey dots) and the satellite (the grey open circle at the center of the figure) are drawn with their actual sizes in the simulation. 
The two ends of the clusters correspond to the Lagrange L1 and L2 points, and the Roche robe shapes the cluster. 
In the left figure (Model 1), the cluster is isolated from the gap edges, while in the right figure (Model 11), it is connected to the gap edges because of the strong gravitational attraction of particles by the heavy satellite.
Particles in the cluster are constantly replaced through the L1 and L2 points. 
The clustering of particles was also observed in the previous local N-body simulation of the formation of propellers by \cite{Lewis:2009}. 
Our simulation suggests that the cluster also forms during early gap opening in the case of a fully opened gap.

This result could be related to the formation history of the small embedded satellites such as Pan and Daphnis. 
It is observationally suggested that Pan and Atlas have the equatorial ridges \citep{Thomas:2019, Buratti:2019} and the ridge could have been formed during the accumulation phase of porous ring particles onto the massive cores \citep[e.g.,][]{Porco:2007, Charnoz:2007, Yasui:2014, Quillen:2021}. 
On the other hand, the shape of Daphnis is a prolate spheroid and its major axis does not align to the radial direction \citep{Tiscareno:2019}.  
The lemon-shaped cluster obtained in our simulation is different from the ``flying saucer'' shape (with a ridge near the equator) of Pan and Atlas and the prolate spheroid of Daphnis.
\cite{Charnoz:2007} claimed that later evolutionary processes such as material redistribution may form such an observed ridge near the equator. 
An embedded satellite may also gain angular momentum from colliding particles to obtain a certain spin velocity \citep[e.g.,][]{Morishima:2004, Ohtsuki:2004}, which could lead to the redistribution of the clustering materials. 
On the other hand, \cite{Leleu:2018} proposed an another formation scenario. 
They demonstrated with combining N-body and SPH simulations that the head-on merging collisions of a comparable size objects can naturally explain the observed shapes with large equatorial ridges.
To address this issue in more detail, our model needs to be revised, for example, in terms of tangential friction, stickiness, size distribution, and spin rotation of the satellite, which is left for future works.

\section{Conclusions and Future Work}\label{Conclusion}

The Cassini observations have revealed striking ring structures created by an embedded satellite (\Figref{fig:image}). 
Most of the previous studies on the formation of these structures have used 1D diffusion calculations, the streamline model, or local N-body simulation, focusing on one or some of the structure(s). 
Here, we have performed high-resolution ($N=1\times 10^6-5\times 10^6$) global full N-body simulation of systems of rings and an embedded satellite.
In this paper, as a first attempt at the global simulation, we have investigated the gap opening, the satellite wakes excited near the gap edge by the satellite, the angular momentum transfer due to wakes generated by the ring self-gravity, the origin of sharp edges of the gap and the mountain-like vertical wall at the wavefronts of satellite wakes. 
Because we performed a global full N-body simulation, all of the above processes consistently occur and their interplay simultaneously creates the striking ring structures.
Although even with $N=1\times 10^6-5\times 10^6$ the simulated particles and satellite are much larger than those in the real system, most of the mechanisms which are discussed here are scaled with the Hill radius of the satellite, and relative importance of the collisional and the self-gravity wakes viscosities is marginally consistent with that in the real system.
Thus the results here may be applied to the real system.
Our findings are summarized as follows:
\begin{itemize}
    \item The interplay of the satellite wakes excited near the gap edge by the satellite, their dissipation by inelastic collisions, and the angular momentum transfer due to wakes generated by the ring self-gravity, together create the sharp gap edges and the mountain-like vertical wall at the wavefronts of satellite wakes.
    
    \item
    The observed clear gap with the satellite wakes near its edges (\Figref{fig:snap} and \Figref{fig:wake}) is reproduced. 
    The gap edge is extremely sharp compared to that of the gas disk (\Figref{fig:sur_den}). 
    The sharpness is enhanced by excitation of the particle eccentricity by the satellite followed by its collisional damping in high-density gap edge (\Figref{fig:streamline}), which was also shown in the local N-body simulation of \cite{Lewis:2011}.
    
    \item The angular momentum flux reversal is identified in the high-density region where the particle trajectories are concentrated.
    It is consistent with the streamline compression argument in the streamline model \citep[e.g.,][]{Borderies:1989} and the local N-body simulation studies \citep[e.g.,][]{Lewis:2000}. 
    \item The formation mechanism of the vertical wall is quantitatively identified: i) Particle's epicyclic motions are exited by the satellite's perturbation; ii) While the epicyclic motions are synchronized and collisional velocity is low in the early first wave, they deviate from the synchronization to produce high collisional velocity in the second epicycle wakes; iii) The lateral epicycle motions are converted into the vertical motion by a single or a few high-velocity particle-particle collision(s) in the high-density wavefronts, because of its spherical shape.
    This mechanism naturally leads to formation of the mountain-like vertical wall structure associated with the satellite wakes, peaked at the second wakes (\Figref{fig:3Dwake}).
    Because the epicycle amplitude is predicted by the satellite Hill radius ($r_{\rm H,s}$) and the half width of the gap, the wall height is semi-analytically predicted to be $\sim 0.1 \, r_{\rm H,s}$ for Daphnis and $\sim 0.01 \, r_{\rm H,s}$ for Pan.
    These results are consistent with the observed shadow cast on the ring plane \citep{Weiss:2009}. 
\end{itemize}

In this paper, we have fixed the co-planer circular orbit of a satellite as a first attempt at our high-resolution global N-body simulation.
However, Daphnis actually has non-negligible eccentricity $e \simeq 3.31\times10^{-5}\pm0.62\times10^{-5}$ and inclination $i\simeq 0^{\circ}0036\pm0^{\circ}0013$  \citep{Jacobson:2008, Weiss:2009}. 
The corresponding radial and vertical amplitudes of the oscillation are 4.5 km  (0.9$\,r_{\rm H}$) and 8.6 km (1.7$\,r_{\rm H}$), respectively \citep{Jacobson:2008}. 
Actually, the time-varying amplitude of the wakes on the Keeler gap edge was observed, for which the eccentricity could be responsible \citep{Seiss:2010}. 
In addition, the mechanism to maintain the high inclination of Daphnis has not been understood.
In the next paper, we plan to perform the global N-body simulation with a satellite in the eccentric and inclined orbit and address these problems.

The fixed satellite orbit highlights the effects of the satellite's perturbations on rings, by separating from orbital changes of the satellites due to the back-reaction.
However, the satellite's orbit and gap structures should evolve by interacting with each other. 
For example, secular evolution of the ring and embedded satellite is suggested to create the spiral density and bending waves \citep{Hahn:2007, Hahn:2008}. 
In addition, the non-Kelperian radial motion of propeller moonlets was observed \citep{Tiscareno:2010}, which may be explained by the interactions between propeller moonlets and stochastic density fluctuations \citep{Crida:2010, Rein:2010, Pan:2012a}. 
\cite{Bromley:2013} suggested through their semi-analytical model that the deficit of small satellites with 2-24 km size in A-ring could reflect fast ``type III'' migration (\cite{Masset:2003}), although the deficit could also be related to the difference of the Roche limit radius due to the material strength, which has not been observationally constrained.
In subsequent papers, we will investigate these problems with the global N-body simulation without the satellite's orbit fixed.
From an N-body simulation point of view, it is rather easier to unlock the satellite orbit.

\section{Acknowledgments}
We thank the two reviewers and Keiji Ohtsuki for their careful reading of our paper and many insightful comments.
We also thank \texttt{n-body-with-center} developer: Junichiro Makino.
Numerical computations were carried out on Cray XC50 at the Center for Computational Astrophysics, National Astronomical Observatory of Japan.
This research is supported by JSPS Kakenhi grant 21H04512.

\printcredits

\bibliographystyle{cas-model2-names}

\bibliography{reference}

\bio{}
\endbio

\endbio

\section*{Appendix A. Derivation of the gap profile using the model for a gas disk}

\cite{Kanagawa:2017}'s formula was originally derived for the gap profile in a gas disk, but an equivalent formula in a planetary ring can be obtained as follows. 
The surface density of the gap structure by \cite{Kanagawa:2017} is given by
\begin{equation}
\Sigma(r)=
\left\{
\begin{array}{ll}
\Sigma_{\rm min} &  {\rm{for}} \ \  |r-R_{\rm pl}|<\Delta R_1\\
\Sigma_{\rm gap}(r) & {\rm{for}} \ \ \Delta R_1<|r-R_{\rm pl}|<\Delta R_2\\
\Sigma_0 & {\rm{for}} \ \ |r-R_{\rm pl}|>\Delta R_2
\end{array}
\right.
,
\label{eq:Sigma}
\end{equation}
where $R_{\rm pl}$ is the orbital radius of a planet. $\Sigma_{\rm gap}(r), \Delta R_1, \Delta R_2$ is defined as 
\begin{equation}
\frac{\Sigma_{\rm gap}(r)}{\Sigma_0}=4.0K'^{-1/4}\frac{|r-R_{\rm pl}|}{R_{\rm pl}} - 0.32,
\label{eq:Sigma_gap}
\end{equation}

\begin{equation}
\Delta R_1=\qty(\frac{\Sigma_{\rm min}}{4\Sigma_0} + 0.08)K'^{1/4}R_{\rm pl},
\label{eq:R1}
\end{equation}

\begin{equation}
\Delta R_2=0.33K'^{1/4}R_{\rm pl},
\label{eq:R2}
\end{equation}
where $K'$ is defined as 
\begin{equation}
K'=\qty(\frac{M_{\rm pl}}{M_{*}})^2\qty(\frac{h}{R_{\rm pl}})^{-3}\alpha^{-1}.
\label{eq:K'}
\end{equation}
$M_{\rm pl}$ is a mass of planet, $M_*$ is a mass of central star and $h$ is a scale height of the disk. They adopted $\alpha$-prescription (\cite{Shakura:1973}), thus the viscosity is give by $\nu=\alpha h^2\Omega$, where $\Omega$ is the local Kepler angular velocity.
Then, we can rewrite \Eqref{eq:K'} in terms of the viscosity $\nu$:
\begin{equation}
K'=\qty(\frac{M_{\rm pl}}{M_{*}})^2\qty(\frac{h}{R_{\rm pl}})^{-5}\qty(\frac{\nu}{R_{\rm pl}^2\Omega})^{-1}.
\end{equation}
The surface density at the bottom of the gap $\Sigma_{\rm min}$ is given by
\begin{equation}
\frac{\Sigma_{\rm min}}{\Sigma_0}=\frac{1}{1+0.04K},
\end{equation}
where $K$ is defined as 
\begin{align}
K&=\qty(\frac{M_{\rm pl}}{M_*})^2\qty(\frac{h}{R_{\rm pl}})^{-5}\alpha^{-1}\notag\\
&=\qty(\frac{M_{\rm pl}}{M_{*}})^2\qty(\frac{h}{R_{\rm pl}})^{-7}\qty(\frac{\nu}{R_{\rm pl}^2\Omega})^{-1}.
\label{eq:K}
\end{align}

In order to adopt this formula to a planetary ring, the scale height and the viscosity in a planetary ring should be evaluated from some previous N-body simulation studies. Here, we assume that the distribution of ring particles is uniform and the self-gravity wakes do not appear ($Q\lesssim2$). In this case, in a planetary ring, the scale height is $h\sim\sigma_r/\Omega$, then we can obtain the aspect ratio using \Eqref{eq:sigma} \citep[e.g.,][]{Daisaka:1999}:
\begin{equation}
\frac{h}{R}=
\left\{
\begin{array}{ll}
\qty(\frac{m_{\rm p}}{M_{\rm s}})^{1/2}\qty(\frac{R}{r})^{-1/2} &  r_{\rm h}^*\lesssim0.5\\
2\qty(\frac{R}{r}) & r_{\rm h}^*\gtrsim0.5
\end{array}
\right.
,
\label{eq:scale_height}
\end{equation}
where $R$ is the radius of a ring particle. The viscosity is given by \Eqref{eq:viscosity_thin}.
Substituting \Eqref{eq:scale_height} and \Eqref{eq:viscosity_thin} for \Eqref{eq:K'} and \Eqref{eq:K}, then we can gain the gap profile in a planetary ring system through \Eqref{eq:Sigma} to \Eqref{eq:R2}. Note that we should replace the planet mass $M_{\rm pl}$ with the satellite mass $M_{\rm s}$ and mass of central star $M_*$ with Saturn's mass $M_{\rm p}$ in Kanagawa's formula when substituting quantities of a planetary ring.

\section*{Appendix B. Simple numerical test}
\begin{figure}[h]
\centering
\includegraphics[width=0.5\linewidth]{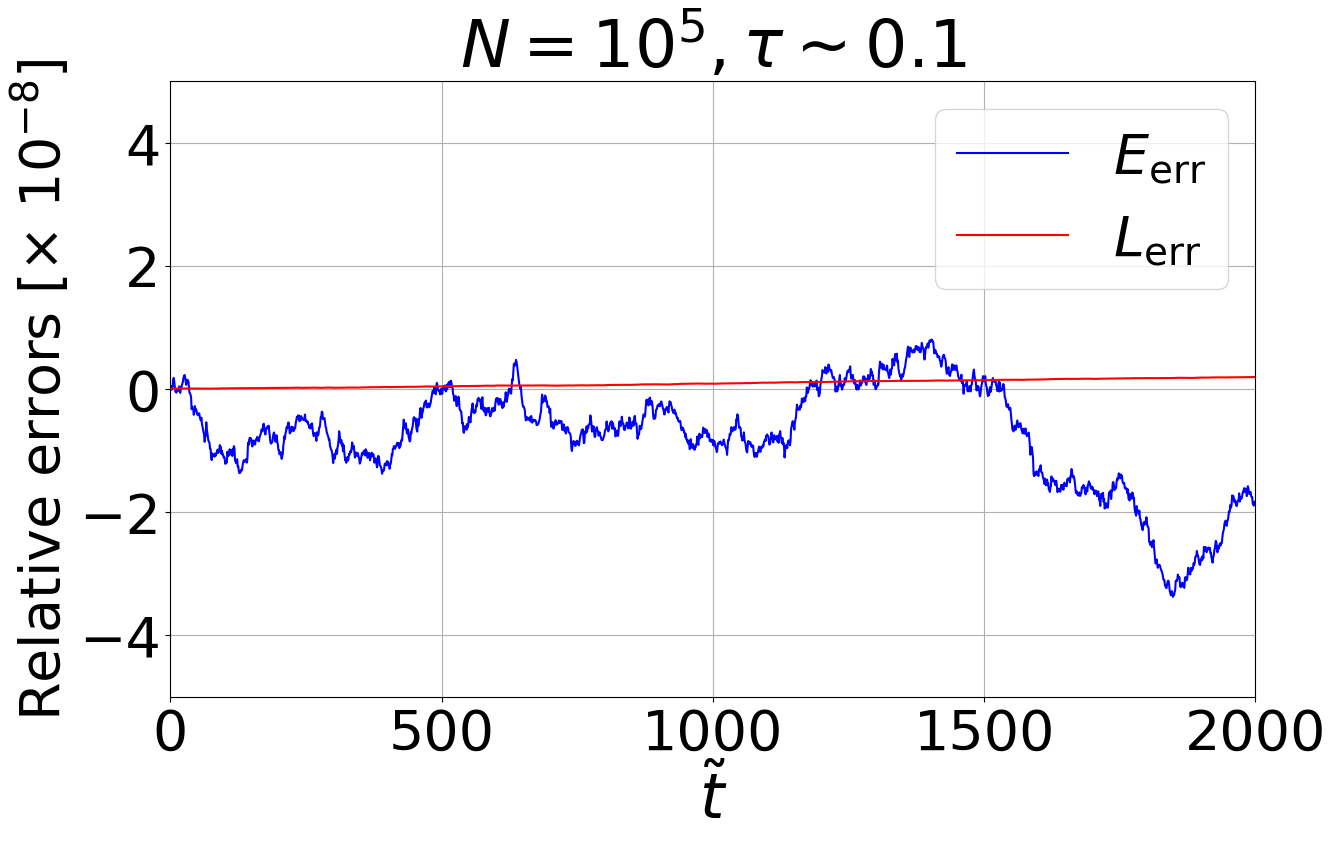}
\caption{The relative errors of total energy $E_{\rm err}$ and angular momentum $L_{\rm err}$ in a test simulation.}
\label{fig:EnergyError}
\end{figure}

Here, we present a simple numerical test for \texttt{n-body-with-center}. 
We have checked the conservation of total energy and angular momentum with $10^5$ particles with making collisions elastic ($\varepsilon=1.0$) and optical depth $\sim 0.01$. 
\Figref{fig:EnergyError} shows the time evolution of total energy and angular momentum relative errors, which are defined as $E_{\rm err}=(E_{\rm now} - E_{\rm ini})/E_{\rm ini}$ and $L_{\rm err}=(L_{\rm now} - L_{\rm ini})/L_{\rm ini}$, where $E_{\rm ini}$ and $L_{\rm ini}$ are the initial total energy and angular momentum, and $E_{\rm now}$ and $L_{\rm now}$ are the current total energy and angular momentum, respectively. Total energy is the sum of the all particles' kinetic energy, gravitational energy and elastic energy during a collision.
The relative error of total energy is within $\sim5\times{10}^{-8}$ and that of total angular momentum is by orders of magnitude better, which is good enough for N-body simulations.

\end{document}